\documentclass[twocolumn]{aastex61}

\usepackage{color}
\usepackage{natbib}
\usepackage{amsmath}
\usepackage{comment}

\newcommand{\ye}{Y_{\rm e}}
\newcommand{\bs}{\boldsymbol}
\newcommand{\pa}{\partial}

\newcommand{\rd}{{\rm d}}

\received{\today}
\revised{\today}
\accepted{\today}

\submitjournal{ApJ}

\shorttitle{Rotating Supernova with Boltzmann $\nu$-Transport}
\shortauthors{Harada et al.}

\begin{document}

\title{On the Neutrino Distributions in Phase Space for the Rotating Core-collapse Supernova Simulated with a Boltzmann-neutrino-radiation-hydrodynamics Code}

\correspondingauthor{Akira Harada}
\email{harada@utap.phys.s.u-tokyo.ac.jp}

\author{Akira Harada}
\affiliation{Physics Department, University of Tokyo, 7-3-1 Hongo, Bunkyo, Tokyo 113-0033, Japan}

\author{Hiroki Nagakura}
\affiliation{Department of Astrophysical Sciences, Princeton University, Princeton, NJ 08544, USA}
\affiliation{TAPIR, Walter Burke institue for Theoretical Physics, Mailcode 350-17, California Institute of Technology, Pasadena, CA 91125, USA}

\author{Wakana Iwakami}
\affiliation{Yukawa Institute for Theoretical Physics, Kyoto University, Oiwake-cho, Kitashirakawa, Sakyo-ku, Kyoto, 606-8502, Japan}
\affiliation{Advanced Research Institute for Science and Engineering, Waseda University, 3-4-1 Okubo, Shinjuku, Tokyo 169-8555, Japan}

\author{Hirotada Okawa}
\affiliation{Yukawa Institute for Theoretical Physics, Kyoto University, Oiwake-cho, Kitashirakawa, Sakyo-ku, Kyoto, 606-8502, Japan}
\affiliation{Advanced Research Institute for Science and Engineering, Waseda University, 3-4-1 Okubo, Shinjuku, Tokyo 169-8555, Japan}

\author{Shun Furusawa}
\affiliation{Interdisciplinary Theoretical and Mathematical Sciences Program (iTHEMS), RIKEN, 2-1 Hirosawa, Wako, Saitama 351-0198, Japan}

\author{Hideo Matsufuru}
\affiliation{High Energy Accelerator Research Organization, 1-1 Oho, Tsukuba, Ibaraki 305-0801, Japan}

\author{Kohsuke Sumiyoshi}
\affiliation{Numazu College of Technology, Ooka 3600, Numazu, Shizuoka 410-8501, Japan}

\author{Shoichi Yamada}
\affiliation{Advanced Research Institute for Science and Engineering, Waseda University, 3-4-1 Okubo, Shinjuku, Tokyo 169-8555, Japan}

\begin{abstract}
With the Boltzmann-radiation-hydrodynamics code, which we have developed to solve numerically the Boltzmann equations for neutrino transfer, the Newtonian hydrodynamics equations, and the Newtonian self-gravity simultaneously and consistently, we simulate the collapse of a rotating core of the progenitor with a zero-age-main-sequence mass of $11.2\,M_\odot$ and a shellular rotation of $1\,{\rm rad\,s^{-1}}$ at the center. We pay particular attention in this paper to the neutrino distribution in phase space, which is affected by the rotation. By solving the Boltzmann equations directly, we can assess the rotation-induced distortion of the angular distribution in momentum space, which gives rise to the rotational component of the neutrino flux. We compare the Eddington tensors calculated both from the raw data and from the M1-closure approximation. We demonstrate that the Eddington tensor is determined by complicated interplays of the fluid velocity and the neutrino interactions and that the M1-closure, which assumes that the Eddington factor is determined by the flux factor, fails to fully capture this aspect, especially in the vicinity of the shock. We find that the error in the Eddington factor reaches $\sim 20\%$ in our simulation. This is due not to the resolution but to the different dependence of the Eddington and flux factors on the angular profile of the neutrino distribution function, and hence modification to the closure relation is needed.

\end{abstract}


\section{Introduction} \label{sec:intro}
The explosion mechanism of core-collapse supernovae (CCSNe) is one of the big issues in astrophysics \citep[][for a review.]{2012ARNPS..62..407J}. The CCSNe are thought to be the explosive death of massive stars and one of the missing pieces of stellar evolution theory. The explosion mechanism is addressed only by numerical simulations, since hydrodynamics is coupled with several complicated physical processes like weak interactions with neutrinos, strong interactions among an ensemble of nuclei, general relativistic gravity, and so on. The CCSNe are the birthplaces of neutron stars, whose merger is currently supposed to be the most promising site for the production of some of the $r$-process elements \citep{2017PhRvL.119p1101A, 2017PASJ...69..102T}, one of the important unknowns in nucleosynthesis theory. In order to understand the history of matter in the universe in a coherent way, unveiling the explosion mechanism of CCSNe is indispensable.

The leading hypothesis for the explosion mechanism is the neutrino heating mechanism \citep{1985nuas.conf..422W}. In this mechanism, the shock wave generated at the core bounce but that stalled thereafter inside the core is re-energized by the absorption of neutrinos emitted from the proto--neutron star (PNS) formed at the center. While spherically symmetric simulations have shown consistently the failure of this mechanism \citep{2001PhRvD..63j3004L, 2005ApJ...629..922S}, multidimensional simulations have emphasized the importance of fluid instabilities such as convection and standing accretion shock instability (SASI) \citep{2012ApJ...761...72M}. These instabilities eventually develop turbulence, which helps the neutrino heating in several ways \citep{2006ApJ...650..291Y, 2012ApJ...749...98T, 2013ApJ...771...52M}. In addition, other physical processes, such as the preexisting turbulence in the outer part of the progenitor \citep{2013ApJ...778L...7C, 2015MNRAS.448.2141M, 2015ApJ...799....5C, 2015ApJ...808L..21C}, and seemingly minor microphysics like the inelastic scattering off nucleons, many-body corrections \citep{2018SSRv..214...33B}, muonic effects \citep{2017PhRvL.119x2702B}, and so on \citep{2018ApJ...853..170K}, have been considered by more recent supernova modelers. The essential ingredient of the explosion mechanism has not been fully understood, though.

In fact, despite a lot of effort devoted to these realistic modelings, there are some puzzles remaining in numerical simulations. First, the explosion energies obtained in the simulations are commonly smaller, just $\sim 1/10$ the typical observed values \citep{2009ApJ...694..664M, 2012ApJ...756...84M, 2014ApJ...786...83T, 2015ApJ...807L..31L, 2015ApJ...801L..24M, 2018SSRv..214...33B, 2016ApJ...825....6S, 2015MNRAS.453..287M, 2018ApJ...854...63O, 2019MNRAS.482..351V}. Longer simulations exceeding several seconds may resolve this problem \citep{2013ApJ...767L...6B, 2016ApJ...818..123B}, but it remains to be demonstrated. Second, the results of simulations are sometimes qualitatively different among groups. This may be partially because these multidimensional simulations employ approximate neutrino transport solvers one way or another, being different from group to group. Since neutrinos are not in equilibrium with matter, their transport should be treated with the Boltzmann equations. Since their numerical solution without imposing spherical symmetry is still highly costly computationally at present, it has been avoided so far (but see \citet{2008ApJ...685.1069O}, in which the authors employed a Boltzmann solver except near the core center, where they adopted an approximation). 
Recently, several works to compare the numerical methods for supernova simulations and to check the influence of the employed approximate methods have been conducted \citep{2016ApJ...831...81S, 2018JPhG...45j4001O, 2018MNRAS.481.4786J, 2018A&A...619A.118C, 2019JPhG...46a4001P, 2018arXiv180910146G}. Such comparisons have just been started, and more works and efforts are required to understand their impact on the CCSN simulations fully.
Especially, in order to calibrate the difference in the approximations, simulations that solve the Boltzmann equations without artificial approximations (other than mandatory finite-differencing of the differential equations) are indispensable.

We have hence developed a Boltzmann-radiation-hydrodynamics code, which solves the Boltzmann equations for neutrino transfer directly by the finite difference without employing any further artificial approximation. 
This code can not only allow us to perform accurate simulations but also play a significant role in the code comparison works.
The basic test of this code was done in \citet{2012ApJS..199...17S} and then \citet{2014ApJS..214...16N} tested the special relativistic extension utilizing the two-grid approach, which is indispensable to treat neutrino trapping in the optically thick region correctly. Finally, \cite{2017ApJS..229...42N} presented the code that incorporated the capability of tracking the proper motion of PNSs and was ready for productive runs of realistic CCSN simulations. The comparison with a Boltzmann solver by the Monte Carlo method was reported in \citet{2017ApJ...847..133R}. The first result produced with this code was reported in \citet{2018ApJ...854..136N}, in which the effect of different equations of state (EOSs) was discussed. Note that the severe limitation of computational resources forces us to impose axisymmetry in our simulations at the moment, although we have already implemented the capability of 3D computations in the code.

In this paper we pay our attention to rotation. As demonstrated in \citet{2018ApJ...854..136N}, only the Boltzmann solver like ours can provide the angular distribution function in momentum space. Note that in the spatially axisymmetric, nonrotating case, the angular distribution in momentum space still has a reflective symmetry with respect to the meridional plane. This symmetry is broken for the rotating system even in the spatial axisymmetry. Detailed examination of such systems will give us a new and deeper insight into the neutrino distributions in the CCSNe. In this paper we assume a modest rotation with which not the dynamics of fluid but the neutrino distributions are affected. It may be true that more exotic features will show up for more rapid rotations, but the dynamics of core collapse and bounce themselves will also be severely modified then, leading, for instance, to the centrifugal bounce. \citep[e.g.,][]{2004ApJ...600..834O}.

This paper is organized as follows: we briefly describe the numerical modeling such as the basic equations to be solved and the progenitor model in section \ref{sec:method}; the shock evolution and other hydrodynamic features are displayed in section \ref{sec:evolution}; the neutrino distributions are discussed in section \ref{sec:distribution}; finally, our findings are summarized in section \ref{sec:summary}. In appendix \ref{sec:diagnostics}, we provide additional information on some diagnostics related to the effects of the rotation.
Unless otherwise stated, we use in equations the unit with $c=G=\hbar=1$ with $c$, $G$, and $\hbar$ being the light speed, the gravitational constant, and the reduced Planck constant, respectively. The metric signature is $-+++$. Greek and Latin indices run over $0\text{--}3$ and $1\text{--}3$, respectively.

\section{Numerical Modeling} \label{sec:method}

We adopt the Boltzmann-radiation-hydrodynamics code based on the discrete-ordinate ($S_N$) method, in which the Boltzmann equation given in the seven-dimensional (one for time, three for space, and another three for momentum) extended phase space is directly discretized. Since the details of the code are explained in \citet{2012ApJS..199...17S, 2014ApJS..214...16N, 2017ApJS..229...42N}, we briefly review only some fundamentals.

The Boltzmann equation is cast into the conservative form in the $(3+1)$-decomposed spacetime \citep[see][for details]{PhysRevD.89.084073}: 
\begin{eqnarray}
\frac{1}{\sqrt{-g}}\frac{\partial}{\partial x^\alpha}\Bigg|_{q_i} \left[ \left( e_{(0)}^\alpha + \sum_{i=1}^3 \ell_{(i)} e_i^\alpha \right) \sqrt{-g}f\right] && \nonumber\\
- \frac{1}{\epsilon^2}\frac{\partial}{\partial \epsilon}(\epsilon^3 f \omega_{(0)}) + \frac{1}{\sin \theta_\nu} \frac{\partial}{\partial \theta_\nu}(\sin \theta_\nu f \omega_{(\theta_\nu)} ) && \nonumber\\
+ \frac{1}{\sin^2 \theta_\nu}\frac{\partial}{\partial \phi_\nu}(f \omega_{(\phi_\nu)}) = S_{\rm rad},
\end{eqnarray}
where $x^\alpha$, and $\epsilon$, $\theta_\nu$, $\phi_\nu$ are the coordinates in the spacetime and momentum space, respectively, and $f$ is the neutrino distribution function; $g$, $e_{(\mu)}^\alpha\,(\mu = 0,\,1,\,2,\,3)$, and $\ell_{(i)}$ are the determinant of the spacetime metric, a set of the local orthonormal tetrad bases, and the directional cosines for the neutrino-propagation direction with respect to $e_{(i)}^\alpha$, respectively. The directional cosines are expressed as $\ell_{(1)} = \cos \theta_\nu$, $\ell_{(2)} = \sin\theta_\nu \cos\phi_\nu$, and $\ell_{(3)} = \sin\theta_\nu\sin\phi_\nu$. The neutrino energy is written as $\epsilon := - p_\alpha e_{(0)}^\alpha$ with the tetrad and the four-momentum of the neutrino $p^\alpha$. The tetrad bases are chosen as follows: we choose the unit vector normal to the spatial hypersurface $n^\alpha$ as $e_{(0)}^\alpha$ and other spatial bases are set to be
\begin{eqnarray}
e_{(1)}^\alpha &=& \gamma_{rr}^{-1/2} \pa_r, \label{eq:tet1} \\
e_{(2)}^\alpha &=& -\frac{\gamma_{r\theta}}{\sqrt{\gamma_{rr}(\gamma_{rr}\gamma_{\theta\theta}-\gamma_{r\theta}^2)}}\pa_r + \sqrt{\frac{\gamma_{rr}}{\gamma_{rr}\gamma_{\theta\theta}-\gamma_{r\theta}^2}}\pa_\theta, \label{eq:tet2} \\
e_{(3)}^\alpha &=& \frac{\gamma^{r\phi}}{\sqrt{\gamma^{\phi\phi}}}\pa_r+\frac{\gamma^{\theta\phi}}{\sqrt{\gamma^{\phi\phi}}}\pa_\theta+\sqrt{\gamma^{\phi\phi}}\pa_\phi, \label{eq:tet3}
\end{eqnarray}
where $\gamma_{ij}$ is the spatial metric for polar coordinates $(r,\theta,\phi)$. The coordinate bases are denoted by $\pa_r$, $\pa_\theta$, and $\pa_\phi$ as usual. In this paper, neutrinos are assumed to be massless. The factors $\omega_{(0)}$, $\omega_{(\theta_\nu)}$, and $\omega_{(\phi_\nu)}$ are given as
\begin{eqnarray}
\omega_{(0)} &:=& \epsilon^{-2}p^\alpha p_\beta \nabla_\alpha e_{(0)}^\beta, \\
\omega_{(\theta_\nu)} &:=& \sum_{i=1}^3 \omega_i \frac{\partial \ell_{(i)}}{\partial \theta_\nu}, \\
\omega_{(\phi_\nu)} &:=& \sum_{i=2}^3 \omega_i \frac{\partial \ell_{(i)}}{\partial \phi_\nu},
\end{eqnarray}
with
\begin{equation}
\omega_i := \epsilon^{-2} p^\alpha p_\beta \nabla_\alpha e_{(i)}^\beta.
\end{equation}
The collision term on the right-hand side is written as $S_{\rm rad}$. Although we use this general relativistic expression for the Boltzmann equation and the code has the capability to solve them, we take into account in this paper only the special relativistic effects: the spatial hypersurface is assumed to be flat, i.e., $\gamma_{ij} = {\rm diag}(1,r^2,r^2\sin^2\theta)$; the lapse function $\alpha$ is set to unity and the shift vector chosen to track the proper motion of the PNS. Note that in this approximation $\sqrt{-g} = r^2\sin\theta$. In order to evaluate the advection terms, we use a combination of the upwind and central difference schemes according to the mean free path. The equations are solved semi-implicitly, and the Bi-CGSTAB method \citep{Saad2003} with the point-Jacobi preconditioner is used for the matrix inversion.

For the hydrodynamics part, we solve the Newtonian equations on the spherical coordinates:
\begin{equation}
\pa_t (\sqrt{-g}\rho) + \pa_i(\sqrt{-g}\rho v^i) = 0,
\end{equation}
\begin{eqnarray}
&&\pa_t (\sqrt{-g}\rho v_r) + \pa_i\left(\sqrt{-g}(\rho v_r v^i + p\delta^i_r)\right) = \nonumber \\
&& \sqrt{-g}\rho \left(-\pa_r \psi + r (v^\theta)^2 + r\sin^2 \theta (v^\phi)^2 + \frac{2p}{r\rho} \right) \nonumber \\
&&-\sqrt{-g}G_r + \sqrt{-g} \rho \dot{\beta_r},
\end{eqnarray}
\begin{eqnarray}
&&\pa_t (\sqrt{-g}\rho v_\theta) + \pa_i\left(\sqrt{-g}(\rho v_\theta v^i + p\delta^i_\theta)\right) = \nonumber \\
&& \sqrt{-g}\rho \left(-r^2 \pa_\theta \psi + \sin\theta \cos\theta (v^\phi)^2 + \frac{p\cos\theta}{\rho\sin\theta}\right) \nonumber \\
&& -\sqrt{-g}G_\theta + \sqrt{-g}\rho \dot{\beta_\theta},
\end{eqnarray}
\begin{eqnarray}
&&\pa_t (\sqrt{-g}\rho v_\phi) + \pa_i\left(\sqrt{-g}(\rho v_\phi v^i + p\delta^i_\phi)\right) = \nonumber \\
&&-\sqrt{-g}\rho\pa_\phi \psi - \sqrt{-g}G_\phi + \sqrt{-g}\rho \dot{\beta_\phi},
\end{eqnarray}
\begin{eqnarray}
&&\pa_t \left(\sqrt{-g} (e + \frac{1}{2}\rho v^2)\right) + \pa_i\left(\sqrt{-g}(e + p + \frac{1}{2}\rho v^2)v^i \right) = \nonumber \\
&& -\sqrt{-g}\rho v^j \pa_j \psi -\sqrt{-g}G_t + \sqrt{-g}\rho v^j \dot{\beta_j},
\end{eqnarray}
and
\begin{equation}
\pa_t(\sqrt{-g}\rho Y_{\rm e} ) + \pa_i (\sqrt{-g}\rho Y_{\rm e}v^i) = -\sqrt{-g}(\Gamma_{\nu_{\rm e}} - \Gamma_{\bar{\nu_{\rm e}}}).
\end{equation}
Here $\rho$, $v^i$, $p$, $e$, $Y_{\rm e}$, $\psi$, and $\beta_i$ are the density, the velocity, the pressure, the internal energy, the electron fraction, the gravitational potential, and the shift vector, respectively. The energy--momentum transfer between neutrinos and matter is given as
\begin{equation}
G^\mu = \int p^\mu \epsilon S_{\rm rad} \rd V_p, \label{eq:interaction}
\end{equation}
where $\rd V_p$ is the invariant volume element in the momentum space. The variation of the electron fraction per unit time that is induced by the emission or absorption of $\nu_{\rm e}$ or $\bar{\nu_{\rm e}}$ is denoted by $\Gamma_i$ ($i=\nu_{\rm e}$ for electron-type neutrinos and $i=\bar{\nu_{\rm e}}$ for anti-electron-type neutrinos) and given as
\begin{equation}
\Gamma_i = m_{\rm u} \int \epsilon S_{{\rm rad},i} \rd V_p,
\end{equation}
with $m_{\rm u}$ and $S_{{\rm rad},i}$ being the atomic mass unit and the corresponding collision term for neutrino species $i$, respectively. The numerical flux is calculated in the HLL scheme \citep{1983SIAMR...25...35H} with the piecewise-parabolic interpolation \citep{Colella1984174}, and the time integration is performed with the second-order Runge--Kutta method. For the gravitational potential $\psi$, we solve the Poisson equation
\begin{equation}
\Delta \psi = 4\pi \rho
\end{equation}
directly. The inverse matrix is constructed by the MICCG method \citep{2011ApJ...731...80N}.

For the comparison of our rotating model with the nonrotating model presented in \cite{2018ApJ...854..136N}, we employ the same progenitor model, i.e., the nonrotating $11.2\,M_\odot$ model taken from \citet{2002RvMP...74.1015W}. We adopt Furusawa's (Furusawa-Shen: FS) multi-nuclear-species EOS \citep{2011ApJ...738..178F, 2013ApJ...772...95F}, which is based on the relativistic mean field theory and also incorporates light nuclei.
The neutrino reactions considered are the same as those in \cite{2018ApJ...854..136N}, being based on the standard set of \citet{1985ApJS...58..771B} but updated in the electron-capture rate by heavy nuclei according to \citet{2010NuPhA.848..454J, 2000NuPhA.673..481L, 2003PhRvL..90x1102L}; the nonelastic scattering off electrons and the nucleon--nucleon bremsstrahlung are also incorporated. Since the neutrino reactions involving $\nu_\mu$, $\bar{\nu_\mu}$, $\nu_\tau$, and $\bar{\nu_\tau}$ are almost the same (but see \cite{2017PhRvL.119x2702B}), these heavy-lepton-type neutrinos are collectively treated and denoted as $\nu_x$. We hence consider three neutrino species of $\nu_{\rm e}$, $\bar{\nu_{\rm e}}$, and $\nu_x$. Although the progenitor is nonrotating originally, we add rotation by hand at the onset of the collapse. The functional form of the rotational velocity is shellular,
\begin{equation}
v^\phi = \frac{1\,{\rm rad/s}}{1+(r/10^8\,{\rm cm})^2}, \label{eq:rot}
\end{equation}
where $r$ is the distance not from the rotational axis but from the center. According to \citet{2015ApJ...811...86Y}, who claim that the progenitor rotation can be detected if the arrival of gravitational waves is observed earlier than the neutronization burst, the rotational velocity in equation (\ref{eq:rot}) is basically too slow to be detected.

The radial mesh covers the region extending from the center to $5000\,{\rm km}$ and divided into 384 bins. The entire meridian section is initially divided into $64$ angular bins. When a negative entropy gradient starts to develop after core bounce, the $\theta$-grid number is doubled to $128$ and we perturb the radial velocity randomly by $0.1\%$ in the region of $30\le r \le 50\,{\rm km}$ artificially as a seed of fluid instabilities. Note that this is the same prescription as in \cite{2018ApJ...854..136N}. As for momentum space, we divide the energy range up to $300\,{\rm MeV}$ into $20$ grid points and the whole solid angle into $10\;(\theta_\nu) \times 6\;(\phi_\nu)$ angular bins. By using K-computer in Riken, whose computational performance is $128\rm GFLOPS\,per\,node$, the simulation of the post-bounce dynamics presented in the following required $1,300,000$ node-hours with $1,536$ nodes and eight cores per node.

\section{The Time Evolution} \label{sec:evolution}

\begin{figure*}[tbph]
\centering
\includegraphics[width=\hsize]{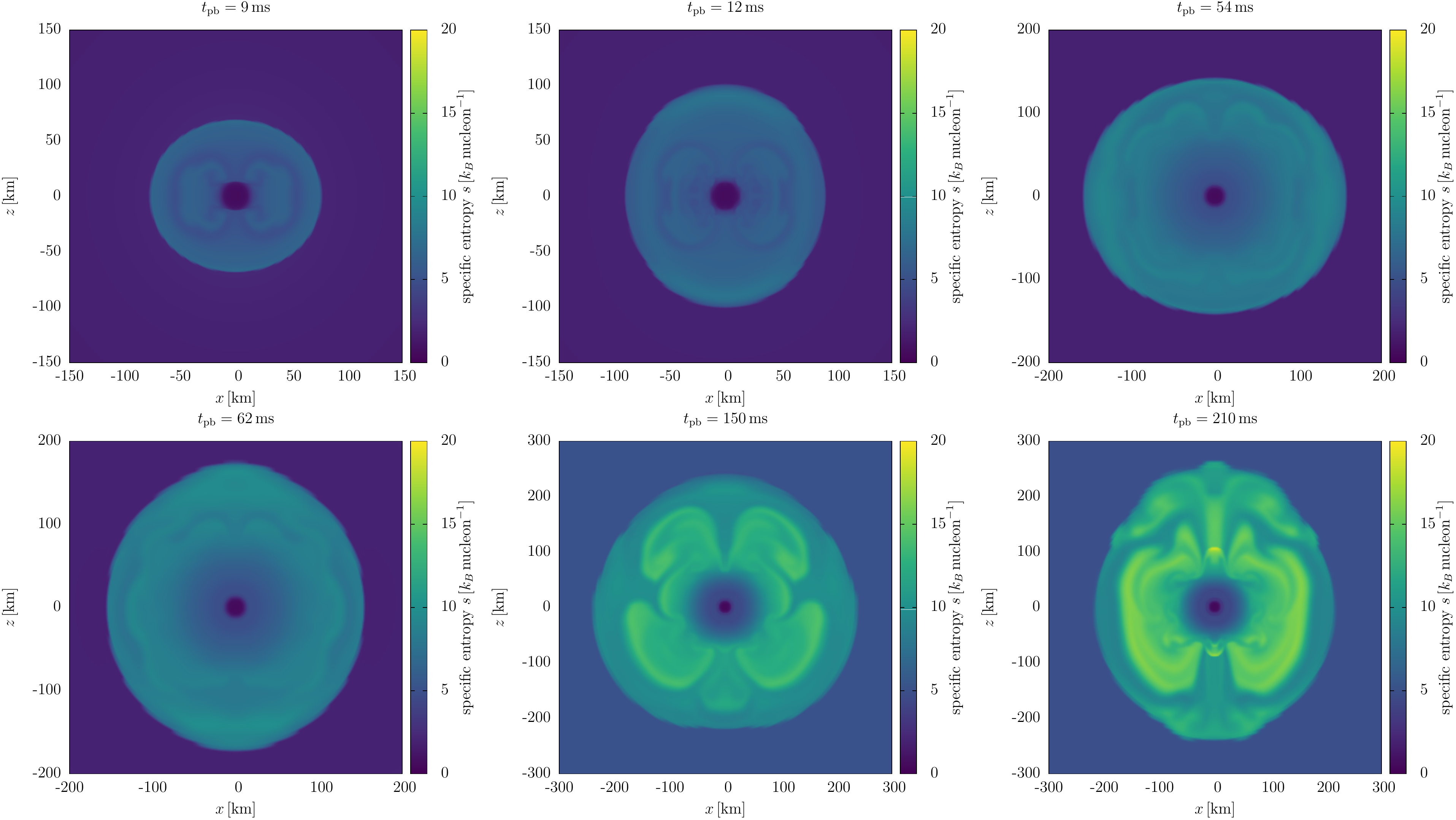}
\caption{\label{fig:snapshots} Entropy distributions in the meridional section at the post-bounce times of $t_{\rm pb} = 9$ (top left), $12$ (top middle), $54$ (top right), $62$ (bottom left), $150$ (bottom middle), and $210\,{\rm ms}$ (bottom right). The colors show the specific entropy whose scale is displayed on the right of each panel. The shock is located at the boundary of the bluish and greenish colors. Note that the ranges of $x$ and $z$ coordinates are different as presented in each figure, indicating the expansion of the shock.}
\end{figure*}

In this section, we give an overview of our simulation by showing several diagnostics for the post-bounce dynamics. First, we display the snapshots of the entropy distributions in figure \ref{fig:snapshots}. They are obtained in the acceleration frame, which moves with the center of PNSs. As shown in appendix \ref{sec:diagnostics}, however, the difference between the laboratory frame and the acceleration frame is very small in this particular model. Thus, we ignore it and call the acceleration frame ``the laboratory frame" hereafter unless otherwise stated.

Right after bounce, the shock expands preferentially in the equatorial direction and takes an oblate shape (top left panel), due to centrifugal force. The radially directed accretion flow is then refracted by the oblate shock in the polar direction. Since axisymmetry is imposed, the refracted flow converges to the rotation axis and is redirected outward, pushing the shock. The shock becomes prolate (bottom left panel). The accretion flow refracted by the prolate shock converges to the equator, pushing the shock equatorially. By repeating this motion, the shock oscillates between the oblate and prolate shapes, with the average shock radius being gradually increased (top right and bottom left panels, respectively). Note that the $\ell=2$ mode deformation of the shock is also observed in \cite{2010PASJ...62L..49S}. In the stalled-shock phase, this oscillation is replaced by the development of convective bubbles. These bubbles have large scales comparable to the scale height and are roughly divided into the northern and southern parts (bottom middle panel). These features are eventually mixed, and a complicated turbulent pattern emerges (bottom right panel).

The flow pattern in the nonrotating model in \citet{2018ApJ...854..136N} is different, on the other hand. Since the centrifugal force is absent in the nonrotating model, the oblate--prolate oscillation seen in figure \ref{fig:snapshots} does not exist. Instead, a rather stochastic pattern presents. Finally, a stochastic turbulent pattern that originated from the convection develops.

\begin{figure}[tbph]
\centering
\includegraphics[width=\hsize]{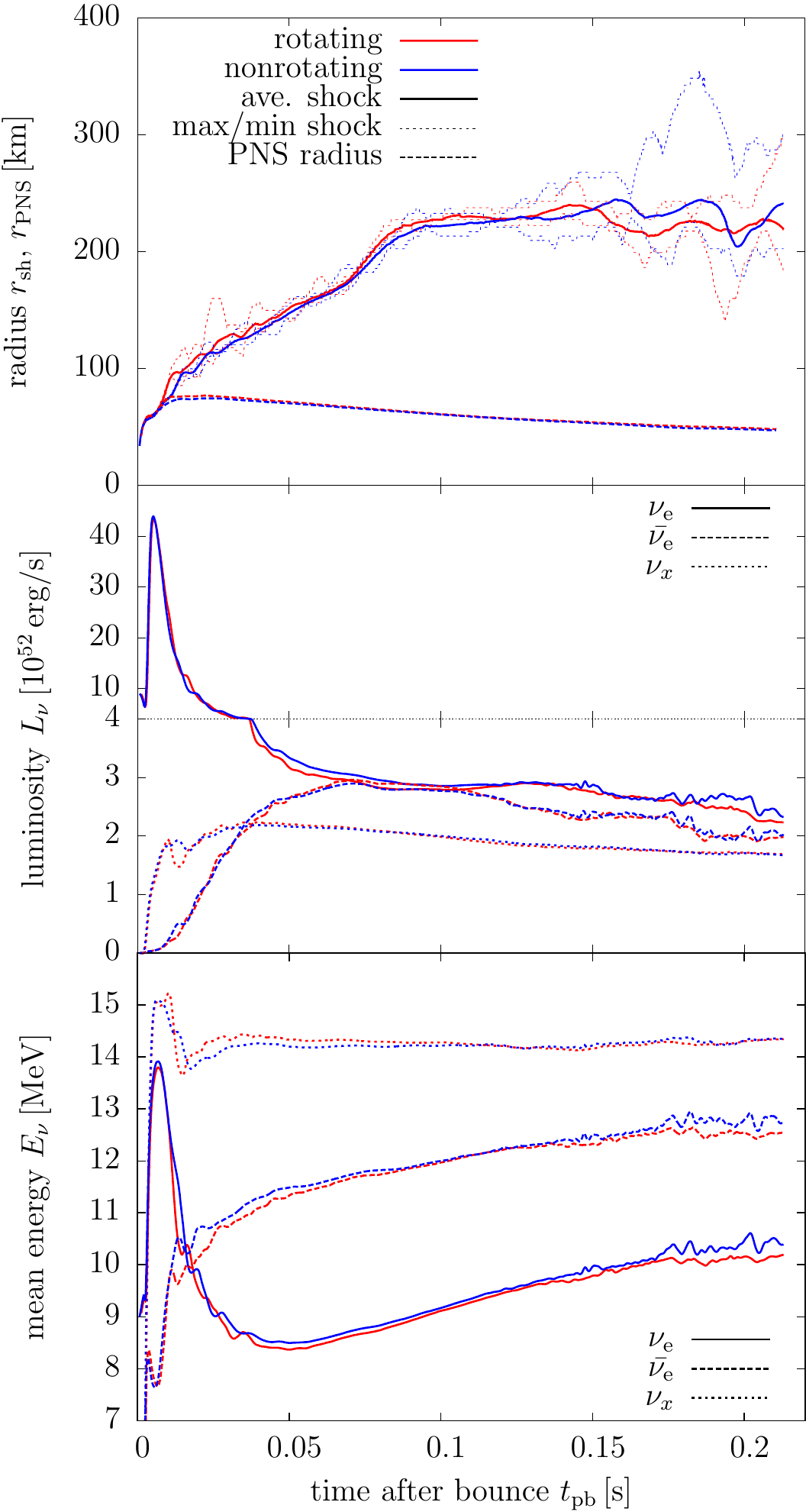}
\caption{\label{fig:hydroevo} Evolutions of some radii and neutrino quantities. For all panels, red and blue lines represent the rotational and nonrotational models, respectively. The top panel shows the shock radii and the PNS radii. The thick solid and thin dotted lines show the average and maximum/minimum shock radii, respectively. The thick dashed lines indicate the PNS radii. The PNS radii are smoothed by the running average over $5\,{\rm ms}$. The middle panel displays the neutrino luminosities. The solid, dashed, and dotted lines correspond to the luminosities of the electron-type neutrinos, anti-electron-type neutrinos, and heavy-lepton-type neutrinos, respectively. Note that the vertical scales of the upper and lower halves of the panel are different, in order both to indicate the peak luminosities at the neutronization bursts and to compare the luminosities of different species at later times. The bottom panel presents the mean energies of neutrinos. The line types are the same for those in the middle panel. Note that the nonrotating model is taken from \citet{2018ApJ...854..136N}.}
\end{figure}

Next, we compare the evolutions of the shock radii $r_{\rm shock}$, the PNS radii $r_{\rm PNS}$, the neutrino luminosities $L_\nu$, and the mean energy of neutrinos $E_\nu$ between the rotating and nonrotating models in figure \ref{fig:hydroevo}. The nonrotating model is taken from \citet{2018ApJ...854..136N}. The shock radius is defined as the outermost radius where the absolute value of the velocity is less than $30\%$ of the freefall velocity. The PNS radius is defined as the radius at which the angle-averaged density is $10^{11}\,{\rm g\,cm^{-3}}$. The luminosities and mean energies of neutrinos are measured at a radius of $500\,{\rm km}$ from the center.

Although the morphology of the shock in the rotating model is affected by the centrifugal force as shown in figure \ref{fig:snapshots}, the evolution of the average shock radius does not much differ from that in the nonrotating model in \citet{2018ApJ...854..136N}. The luminosities and mean energies of neutrinos also have very similar evolutions in the two models. Note that the luminosity of $\nu_{\rm e}$ and the mean energies of $\nu_{\rm e}$ and $\bar{\nu_{\rm e}}$ are slightly smaller for the rotating model. This trend is consistent with \cite{2018ApJ...852...28S}, whose fast-rotating models show smaller neutrino luminosities and mean energies.

It is likely that whether the shock successfully revives or not is determined when the density discontinuity of the progenitor passes through the shock since the ram pressure of the accretion suddenly drops at that time \citep[e.g.,][]{2016ApJ...816...43S, 2018MNRAS.477.3091V}.
Since the shock of the nonrotating model with the FS EOS in \citet{2018ApJ...854..136N} does not revive when the density discontinuity passes through the shock, it seems that the shock revival of the nonrotating model shown in figure \ref{fig:hydroevo} fails. Although the rotating model in this paper is not simulated until the density discontinuity passes, the similarity illustrated in figure \ref{fig:hydroevo} suggests that the rotating model probably fails explosion as well. Some recent works show much later shock revivals \citep{2016ApJ...825....6S, 2018ApJ...854...63O}, but limited computational resources prevent us from running such a long time simulation.
This is the reason why we terminated our simulation at $\sim 200\,{\rm ms}$ after bounce. Note that the dynamics is not the focus of this paper. Dynamics and properties of neutrino emissions for more rapidly rotating models will be reported in the forthcoming paper.

The similarities in the neutrino luminosities and mean energies are originated from the fact that PNS radii are essentially identical as seen in the top panel of figure \ref{fig:hydroevo}. Due to the centrifugal force, the equatorial radius of the PNS is larger than the polar radius by $\sim 5\%$. This is too small to affect the shock evolution in our model.

\begin{figure}[tbph]
\centering
\includegraphics[width=\hsize]{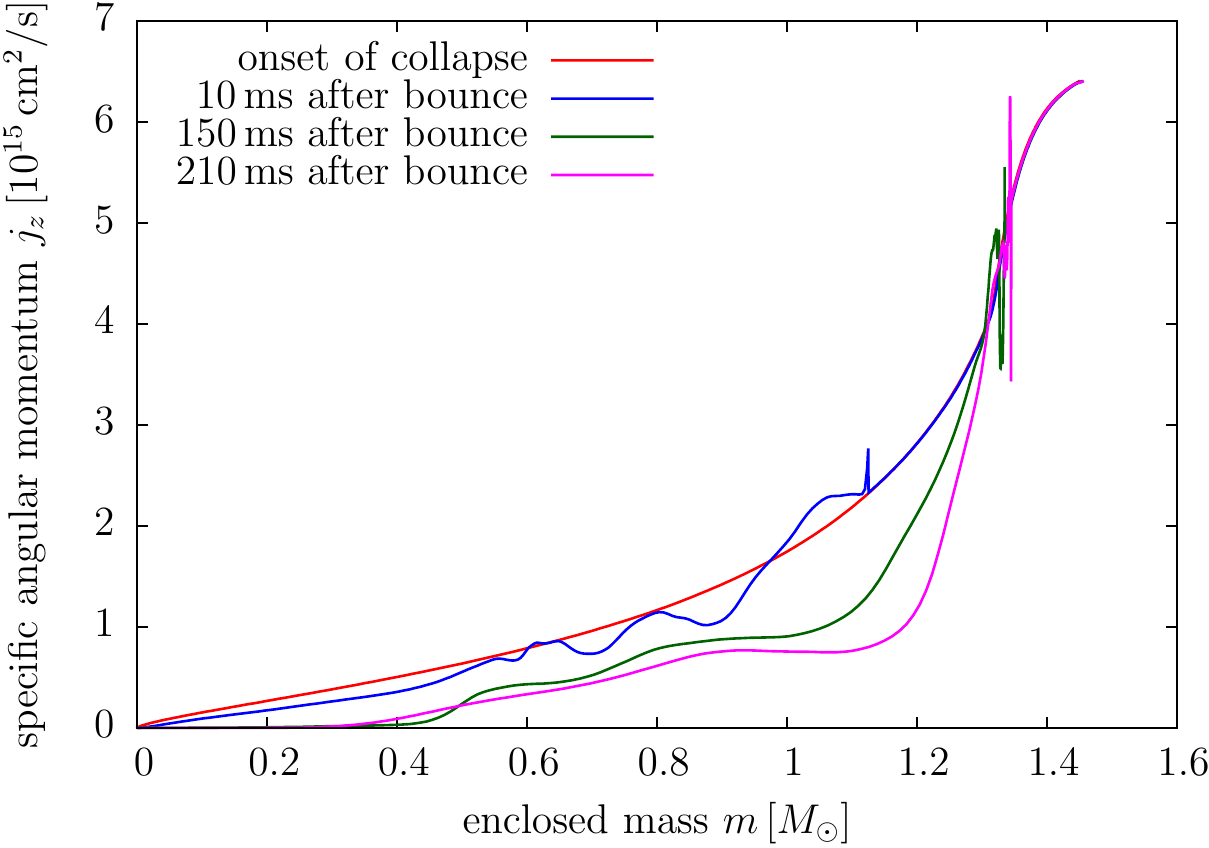}
\caption{\label{fig:angularmomentum} Specific angular momentum of each radial bin as a function of the enclosed mass. Different colors correspond to different times (red: at the onset of the collapse; blue: $\sim 10\,{\rm ms}$ after bounce; green: $\sim 150\,{\rm ms}$ after bounce; magenta: $\sim 210\,{\rm ms}$ after bounce). The spikes in the profiles indicate the positions of the shock at their times.}
\end{figure}

Figure \ref{fig:angularmomentum} presents the evolution of the specific angular momentum defined for radial shells as
\begin{equation}
j_z := \frac{\int_{\rm shell} \rho r^2\sin^2\theta v^\phi \rd V_x}{\int_{\rm shell} \rho \rd V_x},
\end{equation}
where $\rd V_x$ is the invariant volume element in the configuration space and the integration is done over each bin in the radial mesh employed in the simulation. Note that the specific angular momentum inside the shock decreases with time on average. This is because neutrinos carry away some angular momentum (see section \ref{sec:distribution} for a detailed discussion). Although the neutrino emission during the collapse also reduces the angular momentum, it is negligibly small.
In the outer part, where neutrino reactions rarely occur, the angular momentum is essentially conserved. Note that the specific angular momentum distribution in our model lies between the two models ($\sim 10^{14}\,{\rm cm^2\,s^{-1}}$ for the slower model named ``rot'' and $\sim 10^{16}\,{\rm cm^2\,s^{-1}}$ for the faster model named ``artrot'') computed in \citet{2018ApJ...852...28S} although the rotational velocities are higher in our model. This is due to the different progenitor model they employed. The fact that both their ``rot'' model and ours have no influence on the PNS radius whereas their ``artrot'' model did have non-negligible effects may indicate that the border between slow and fast rotations lies between $10^{15}$ and $10^{16}\,{\rm cm^2\,s^{-1}}$.

\section{Neutrino Distribution} \label{sec:distribution}
One of the novel aspects of our code is to treat not the angular moments but the distribution functions of neutrinos directly. In this section we provide detailed analyses of the neutrino distributions.

\subsection{Angular Distribution}

\begin{figure*}[tbph]
\centering
\includegraphics[width=\hsize]{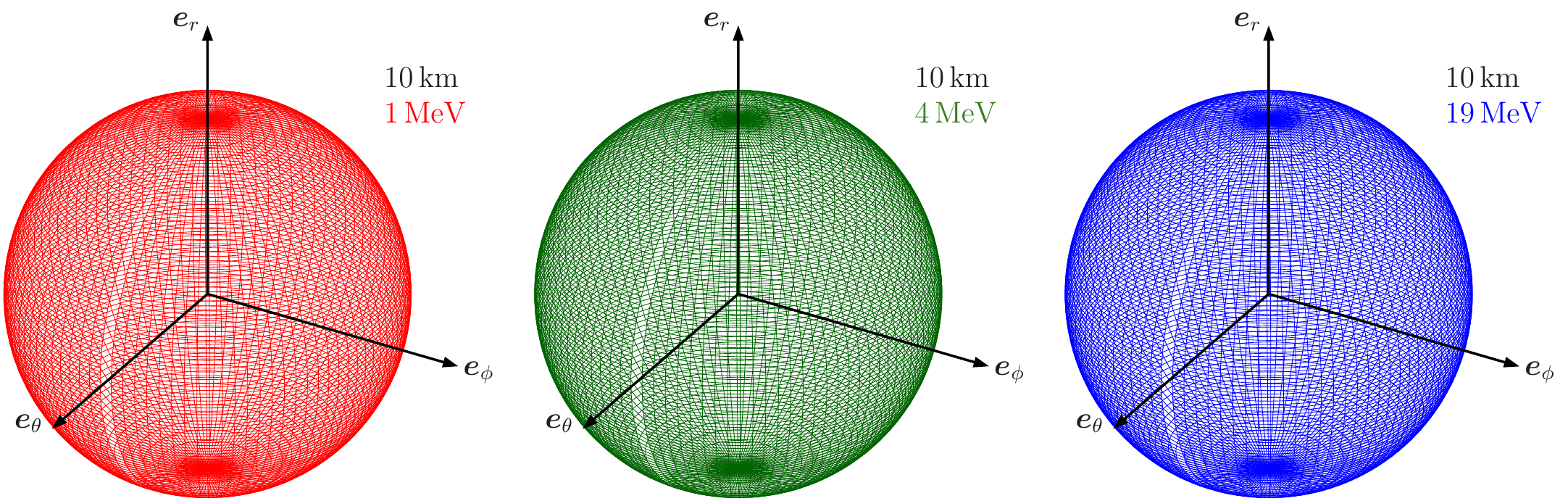}
\caption{\label{fig:distri_thick} Angular distributions in momentum space of the electron-type neutrino at $12\,{\rm ms}$ after bounce in the laboratory frame. The spatial point is $r=10\,{\rm km}$ in the optically thick region on the equator. Each panel represents different neutrino energies measured in the laboratory frame: $1\,{\rm MeV}$ (red), $4\,{\rm MeV}$ (green), and $19\,{\rm MeV}$ (blue). Arrows with ${\bs e}_r$, ${\bs e}_\theta$, and ${\bs e}_\phi$ represent the spatial bases of the tetrad (equations (\ref{eq:tet1}--\ref{eq:tet3})). All distributions are normalized so that the maximum value is the same, say, unity. In order to make the surfaces smooth, angular interpolation is applied.
}
\end{figure*}

\begin{figure*}[tbph]
\centering
\includegraphics[width=\hsize]{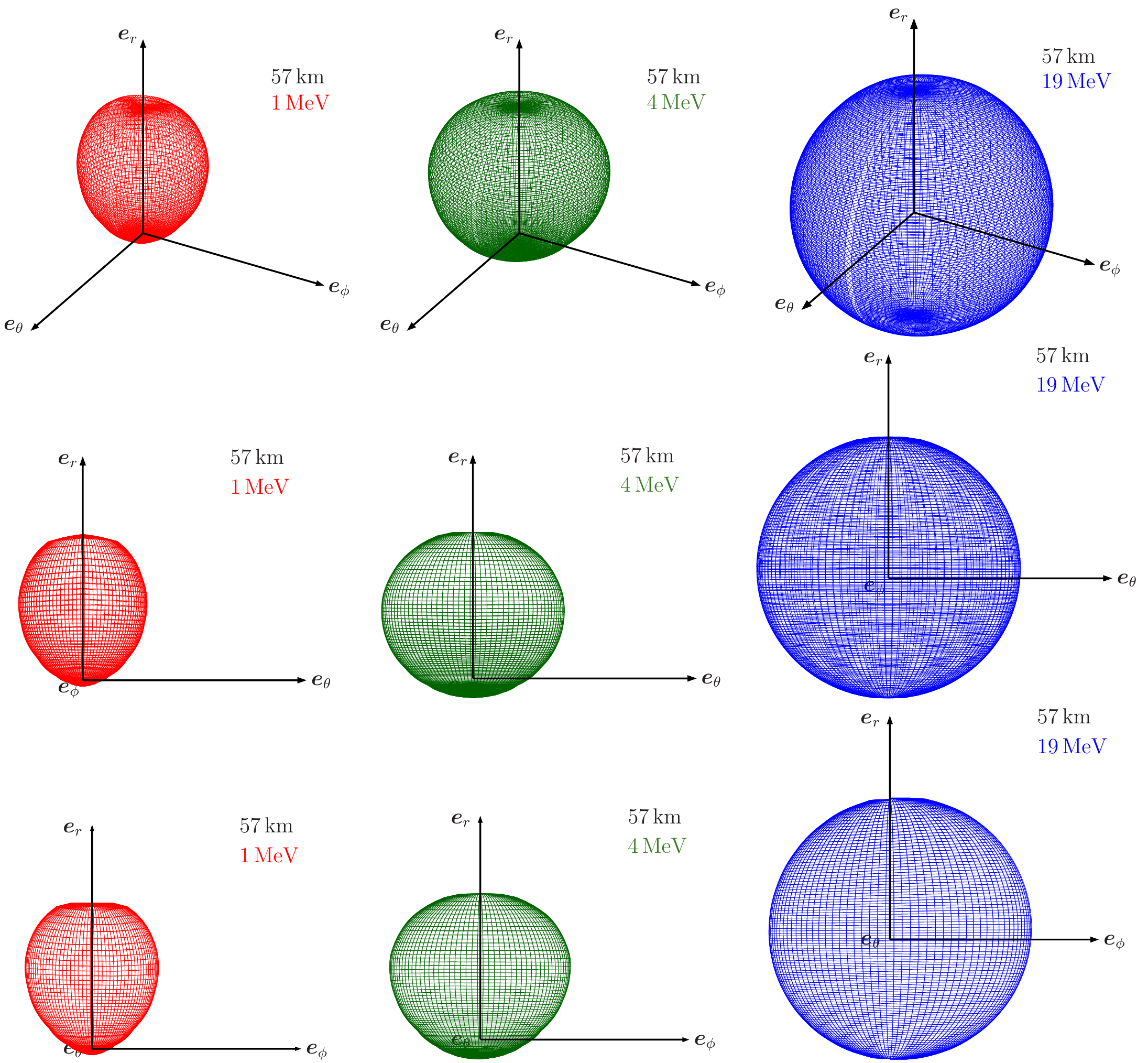}
\caption{\label{fig:distri_semi} Same as figure \ref{fig:distri_thick} except that the spatial point is $r=57\,{\rm km}$ in the semitransparent region. The middle and bottom rows of figures are the distributions projected to $\boldsymbol{e}_r$--$\boldsymbol{e}_\theta$ and $\boldsymbol{e}_r$--$\boldsymbol{e}_\phi$ planes, respectively.}
\end{figure*}

\begin{figure*}[tbph]
\centering
\includegraphics[width=\hsize]{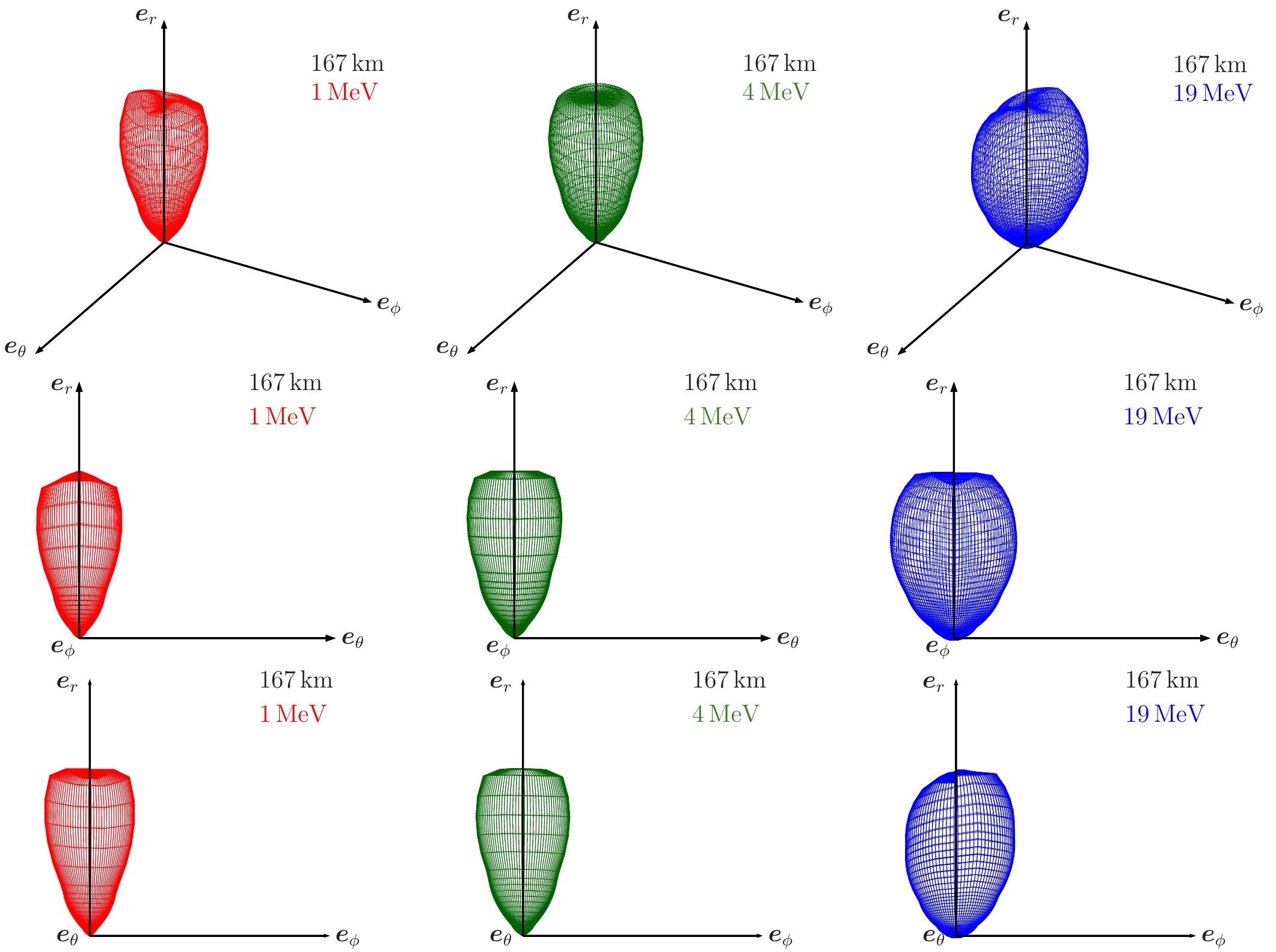}
\caption{\label{fig:distri_thin} Same as figure \ref{fig:distri_semi} except that the spatial point is $r=167\,{\rm km}$ in the optically thin region.}
\end{figure*}

Figures \ref{fig:distri_thick}--\ref{fig:distri_thin} show the angular distributions in momentum space of the electron-type neutrinos with three different energies at $12\,{\rm ms}$ after bounce in the laboratory frame. The spatial locations are chosen from the optically thick (figure \ref{fig:distri_thick}), semitransparent (figure \ref{fig:distri_semi}), and optically thin (figure \ref{fig:distri_thin}) regions, and they are all sitting on the equator ($\theta=\pi/2$). 

In the optically thick region (figure \ref{fig:distri_thick}), neutrinos are in equilibrium with matter and have an isotropic distribution in the fluid rest frame. Since the matter velocity at this point is negligible ($v/c\sim 2\times 10^{-2}$), the distributions are nearly isotropic even in the laboratory frame for all three energies.

On the other hand, the distributions in the semitransparent region (figure \ref{fig:distri_semi}) are obviously not isotropic and are different among three neutrino energies. It is forward-peaked for the lowest-energy neutrinos, while for the middle-energy neutrinos the forward peak is less remarkable. For the highest-energy neutrinos, the distribution is more or less isotropic but slightly elongated in the $\phi$-direction because of the relativistic beaming by the matter rotation. These behaviors are just as expected. Roughly speaking, the neutrino reaction rates are proportional to the squared neutrino energy \citep{1985ApJS...58..771B, 1990RvMP...62..801B, 2001A&A...368..527J}. Since the reaction rates are smaller for lower-energy neutrinos, they decouple from matter deeper in the core at higher densities \citep{2006RPPh...69..971K}, leading to larger deviations from isotropic angular distributions.

Then in the optically thin region (figure \ref{fig:distri_thin}), neutrinos with the three energies all have forward-peaked distributions. This can be easily understood since all neutrinos have already been decoupled from the matter and are streaming freely. The streaming directions are slightly different, though. The principal axes in the distributions of the lowest- and middle-energy neutrinos are almost aligned with the radial direction ($\bs e_r$), whereas for the highest-energy neutrinos the distribution is visibly tilted to the rotational direction ($\bs e_\phi$). This is again understood from the dependence of the reaction rates on the neutrino energy as follows.

\begin{figure}[tbph]
\centering
\includegraphics[width=\hsize]{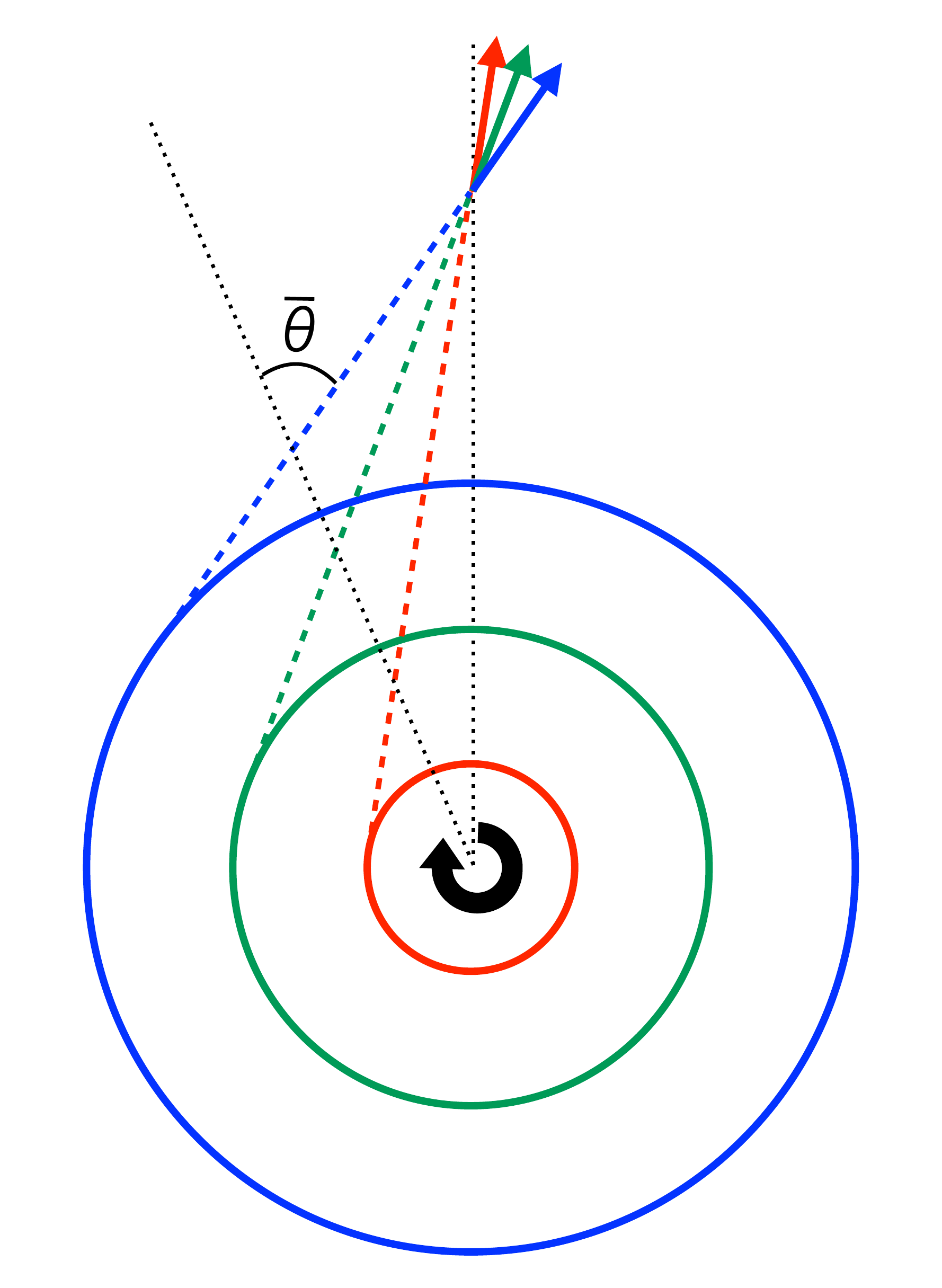}
\caption{\label{fig:tilted_distribution} Schematic picture for the understanding of the neutrino distributions given in figure \ref{fig:distri_thin}. The colored circles represent the equatorial sections of the neutrinospheres for three energies in figure \ref{fig:distri_thin}: $1$, $4$, and $19\,{\rm MeV}$ for red, green, and blue, respectively. The central black circular arrow indicates the rotation of the PNS. The dashed lines and solid arrows are the trajectory of neutrinos and propagating directions, respectively, for three energies. The black dotted lines are drawn along a radial ray in order to emphasize the inclination of the solid arrows. The angle $\bar{\theta}$ in the text is also indicated.}
\end{figure}

The situation is sketched in figure \ref{fig:tilted_distribution}. When neutrinos are trapped by matter, they are dragged by matter and the relativistic beaming occurs, albeit slightly, in the rotational direction as shown with the blue surface in figure \ref{fig:distri_semi}. This tilting remains even after neutrinos are decoupled with matter (see the dashed lines in figure \ref{fig:tilted_distribution}). As neutrinos stream freely to large radii, the angle between the radial and the propagation directions $\bar{\theta}$ gets smaller as $\sin \bar{\theta} = b/r$, where $b$ is the impact parameter with respect to the center. Since the neutrinosphere for higher-energy neutrinos is larger than that for lower-energy neutrinos as discussed by \cite{2006RPPh...69..971K} (compare the blue and red circles in figure \ref{fig:tilted_distribution}),
the impact parameter is larger for the former. As a consequence, the higher the neutrino energy is, the more tilted the distribution is to the $\phi$-direction as shown with the arrows in figure \ref{fig:tilted_distribution}.

\subsection{Rotational Flux}

\begin{figure}[tbph]
\centering
\includegraphics[width=\hsize]{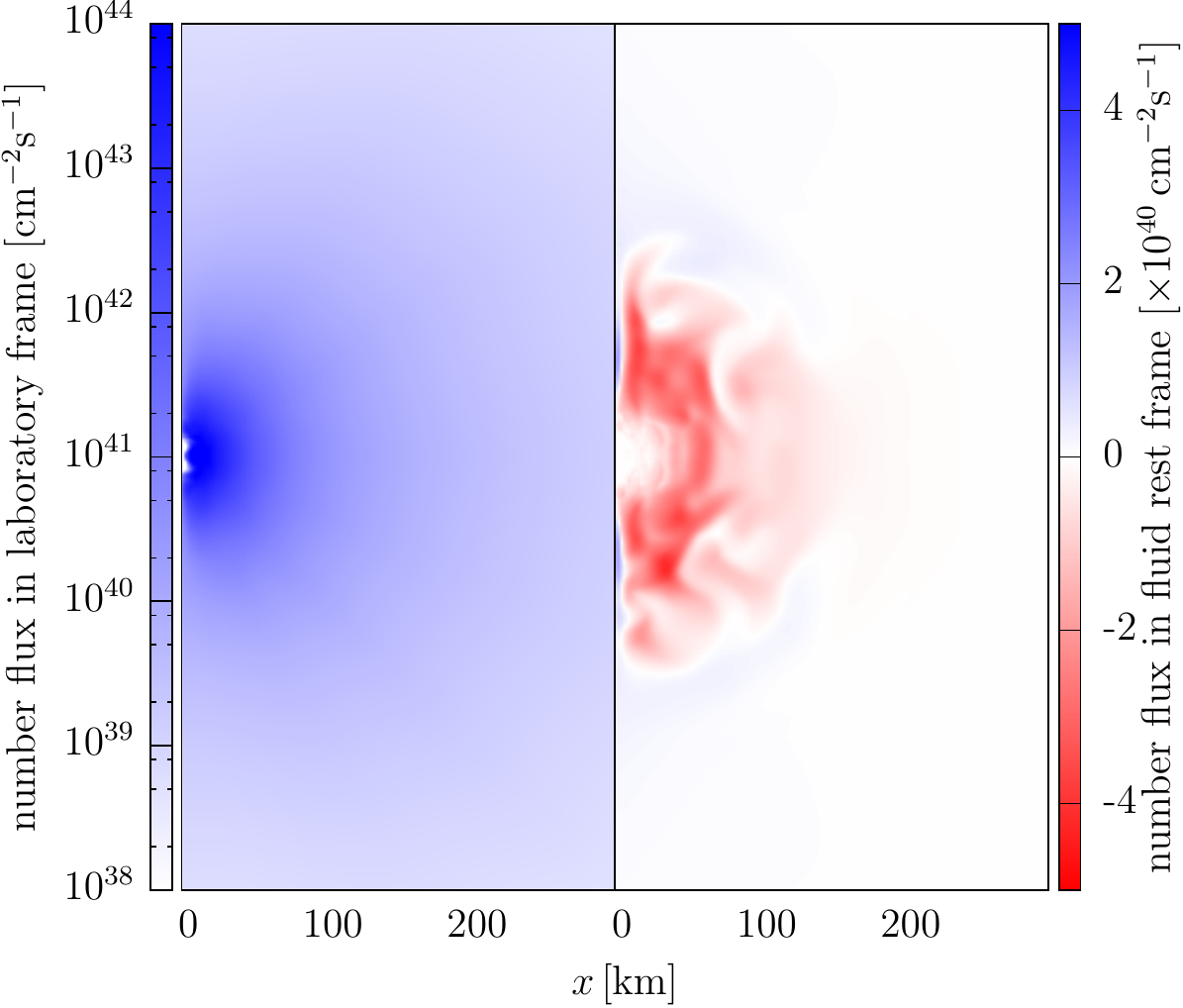}
\caption{\label{fig:flux} Rotational component of the number flux of $\nu_{\rm e}$ at $100\,{\rm ms}$ after bounce in the laboratory frame (left panel) and in the fluid rest frame (right panel). Note that the log scale is used for the left, whereas the linear scale is employed for the right.}
\end{figure}

Since the neutrino distribution is no longer symmetric with respect to the plane spanned by $\bs e_r$ and $\bs e_\theta$ in the presence of rotation, the neutrino flux has a nonzero $\phi$-component in general. This ``rotational'' component is displayed for the electron-type neutrino number flux at $100\,{\rm ms}$ after bounce in figure \ref{fig:flux}. In the left panel, the rotational component measured in the laboratory frame is shown. Since the component is always positive, i.e., neutrinos rotate in the same direction with matter, the log scale is employed in the color bar. This figure demonstrates that the rotational component decreases rapidly with the radius, which is compatible with the above discussion on $\bar{\theta}$.

\begin{figure}[tbph]
\centering
\includegraphics[width=\hsize]{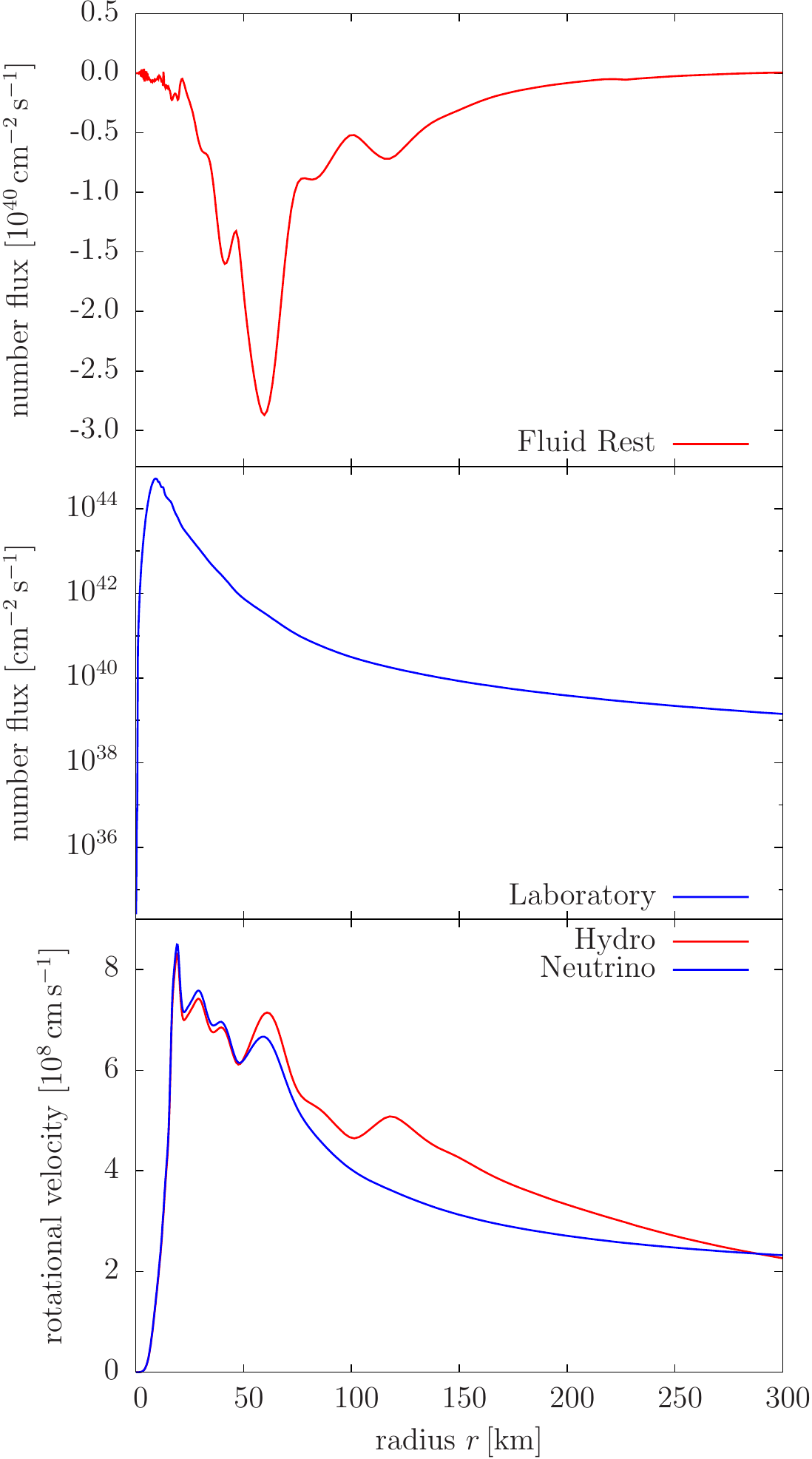}
\caption{\label{fig:alongEWrot} Radial profiles of the rotational component of the number flux of $\nu_{\rm e}$ in the fluid rest frame (top), that in the laboratory frame (middle), and the rotational velocities of matter (bottom, red line) and neutrinos (bottom, blue line), both in the laboratory frame. Note that the middle panel is displayed in the log scale.}
\end{figure}

In the right panel the rotational component in the fluid rest frame is shown. Contrary to the left panel, the color bar is given in the linear scale, since the rotational component can be positive or negative. After the decoupling with matter, the ``rotational velocity'' of neutrinos, which is defined as the $\phi$-component of the number flux divided by the number density of neutrinos, in the laboratory frame declines faster than the rotational velocity of matter. This situation is shown in figure \ref{fig:alongEWrot}, in which radial profiles of the number flux and rotational velocities of matter and neutrinos on the equator are displayed. The $\phi$-component of the fluxes in the fluid rest frame and laboratory frame are again negative and positive, respectively. The rotational velocities of matter and neutrinos in the laboratory frame are almost identical at $r < 50\,{\rm km}$, whereas the matter velocity is larger at larger radii. These results demonstrate that our simulation successfully captures the neutrino transport in the moving matter.

\subsection{Eddington Tensor}

In the often-used Ray-by-Ray(-plus) approximation \citep{2006A&A...447.1049B}, the neutrino distributions are assumed to be axisymmetric with respect to the radial direction. As a consequence, the lateral component of the flux, such as those shown in figure \ref{fig:flux}, is completely neglected. On the other hand, other approximations such as the M1-closure method can treat the nonradially directed flux (\citet{1984JQSRT..31..149L, 2011PThPh.125.1255S}, and see \citet{2012ApJ...755...11K, 2015MNRAS.453.3386J, 2016ApJ...831...81S} for its application to the simulation of CCSNe). As discussed in  \citet{1984JQSRT..31..149L}, the M1-closure method assumes that the neutrino distributions are axisymmetric with respect to the flux and the Eddington factor, which is the largest eigenvalue of the Eddington tensor defined later, is given by a certain prescription. Since our Boltzmann solver does not impose any such artificial assumptions, we can evaluate the validity of these assumptions quantitatively.

As such an attempt, we compare the Eddington tensor calculated according to the definition and that obtained in the M1-closure method. Note that both of them are based on the same numerical data. The Eddington tensor is defined as $k^{ij}(\epsilon) := P^{ij}(\epsilon)/E(\epsilon)$, where
\begin{eqnarray}
P^{ij}(\epsilon) &:=& \gamma^i{}_\sigma \gamma^j{}_\rho M^{\sigma\rho}(\epsilon), \label{eq:boltzpij}\\
E(\epsilon) &:=& n_\sigma n_\rho M^{\sigma\rho}(\epsilon),
\end{eqnarray}
with $M^{\sigma \rho}$ being the second angular moment of the distribution function given as
\begin{eqnarray}
M^{\sigma\rho}(\epsilon) &:=& \int f \delta\left(\frac{\epsilon^3}{3}-\frac{\epsilon^{\prime 3}}{3}\right) p^{\prime \sigma} p^{\prime \rho} \epsilon^\prime \rd \epsilon^\prime \rd \Omega_p^\prime \nonumber \\ &=& \frac{1}{\epsilon} \int f p^{\prime\sigma} p^{\prime\rho} \rd\Omega_p^\prime.
\end{eqnarray}
In this definition, $\Omega_p^\prime$ is the solid angle in the momentum space measured in the fluid rest frame and $\epsilon^\prime \rd \epsilon^\prime \rd \Omega_p^\prime = \rd V_p^\prime$ is the volume element in the same momentum space. Note that this definition is slightly different from that in \citet[see their Equation (2.1)]{2011PThPh.125.1255S}, where they use $\delta (\epsilon - \epsilon^\prime)$ instead of $\delta (\epsilon^3/3 - \epsilon^{\prime 3}/3)$\footnote{\citet{2011PThPh.125.1255S} consider the radiation field in a specific-intensity-like way, and hence the neutrino energy is a natural integral measure. On the other hand, we consider the radiation field as an ensemble of particles, and hence the volume element in momentum space is a natural integral measure.}. This difference does not affect the definition of the Eddington tensor.

In the M1-closure method, on the other hand, the Eddington tensor $k_{\rm M1}^{ij}(\epsilon) := P_{\rm M1}^{ij}(\epsilon)/E(\epsilon)$ is given by the following formula:
\begin{equation}
P_{\rm M1}^{ij}(\epsilon) := \frac{3\zeta(\epsilon)-1}{2}P_{\rm thin}^{ij}(\epsilon)+\frac{3(1-\zeta(\epsilon))}{2}P_{\rm thick}^{ij}(\epsilon). \label{eq:m1pij}
\end{equation}
Here $\zeta(\epsilon)$ is the Eddington factor approximated as \citep{1984JQSRT..31..149L}
\begin{equation}
\zeta(\epsilon) = \frac{3+4\bar{F}(\epsilon)^2}{5+2\sqrt{4-3\bar{F}(\epsilon)^2}}, \label{eq:levermore}
\end{equation}
where $\bar{F}(\epsilon)$ is the flux factor. In this paper, the flux factor is defined in the fluid rest frame as
\begin{equation}
\bar{F}(\epsilon) = \sqrt{\frac{h_{\sigma\rho}H^\sigma(\epsilon)H^\rho(\epsilon)}{J(\epsilon)^2}},
\end{equation}
where 
\begin{equation}
h_{\sigma\rho} := g_{\sigma\rho} + u_\sigma u_\rho
\end{equation}
is the spatial metric projecting onto the fluid rest frame, $u^\sigma$ is the $4$-velocity of matter, and
\begin{eqnarray}
J(\epsilon) &:=& u_\sigma u_\rho M^{\sigma\rho}(\epsilon), \\
H^\sigma (\epsilon) &:=& - h^\sigma{}_\rho u_\lambda M^{\sigma \lambda}(\epsilon)
\end{eqnarray}
are the energy density and energy flux in the fluid rest frame, respectively. In the M1-closure method the optically thin limit $P_{\rm thin}^{ij}(\epsilon)$ and thick limit $P_{\rm thick}^{ij}(\epsilon)$ are smoothly connected. They are defined as
\begin{equation}
P_{\rm thin}^{ij}(\epsilon) = E(\epsilon) \frac{F^i(\epsilon)F^j(\epsilon)}{F(\epsilon)^2}, \label{eq:thinpij}
\end{equation}
and
\begin{equation}
P_{\rm thick}^{ij}(\epsilon) = J(\epsilon) \frac{\gamma^{ij}+4V^i V^j}{3} + H^i (\epsilon) V^j + V^i H^j(\epsilon), \label{eq:thickpij}
\end{equation}
respectively, where we further define
\begin{equation}
F^i(\epsilon) := -\gamma^i{}_\sigma n_\rho M^{\sigma\rho}(\epsilon),
\end{equation}
and $V^i := u^i/u^t$, which is the $3$-velocity of matter. Hereafter we refer to $k^{ij}(\epsilon)$ and $k_{\rm M1}^{ij}(\epsilon)$ as ``the Boltzmann-Eddington tensor'' and ``the M1-Eddington tensor'', respectively. Although one may use the energy-integrated Eddington tensors, we only use the Eddington tensors without energy integration. For the neutrino energy, we adopt the mean energy at each point throughout this section. Note that the M1-Eddington tensor is the same as that used in \cite{2016ApJ...829L..14K} except that a different analytic Eddington factor is adopted. \citet{2015MNRAS.453.3386J} and \citet{2016ApJ...831...81S} employ a similar analytic Eddington tensor while it is defined in the fluid rest frame.

\subsubsection{Physical Interpretation of the Eddington Tensor}

\begin{figure*}[tbph]
\centering
\includegraphics[width=\hsize]{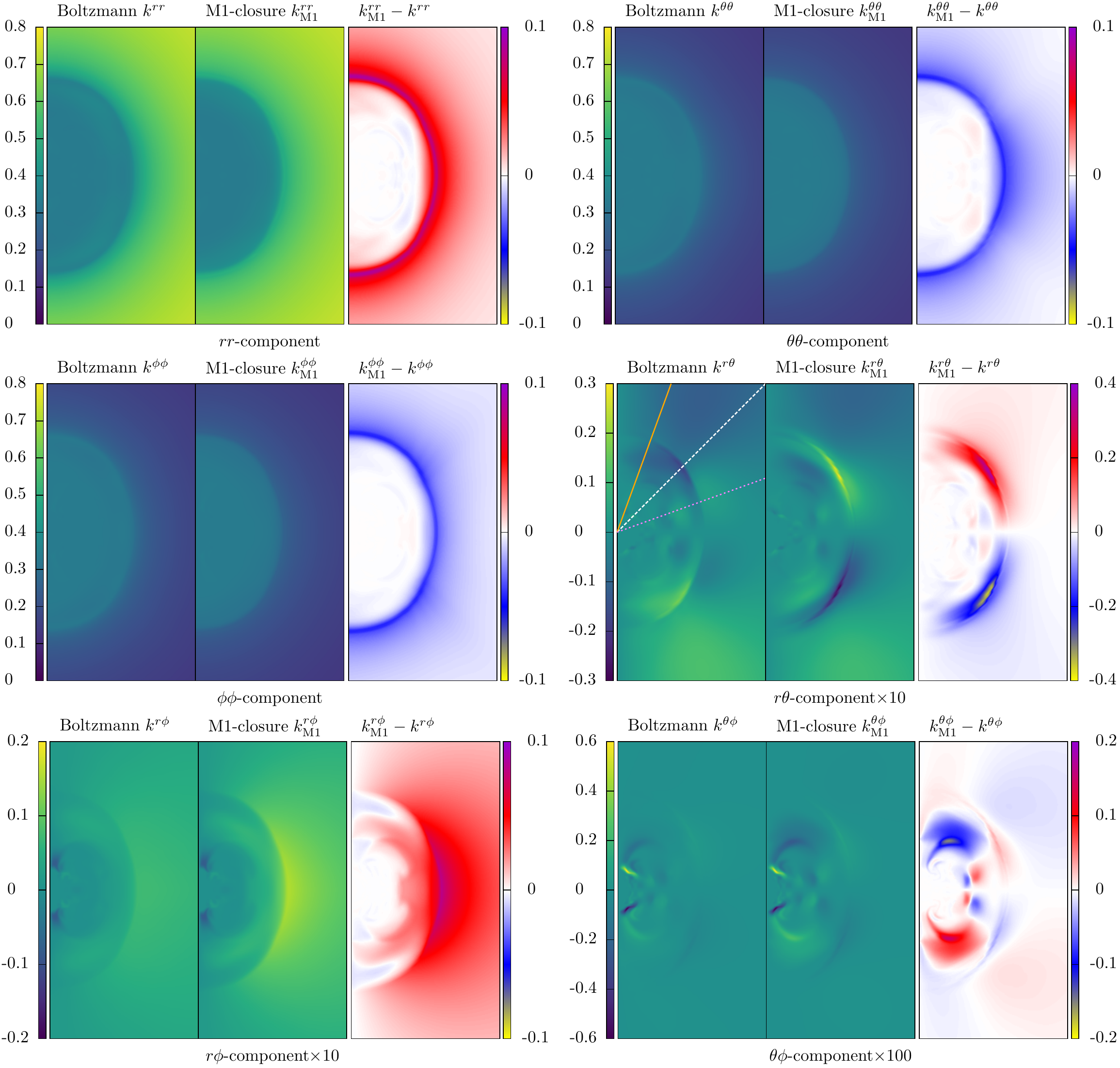}
\caption{\label{fig:eddmap} Comparison of individual components of the Boltzmann- and M1-Eddington tensors, $k^{ij}(\epsilon)$ and $k^{ij}_{\rm M1}(\epsilon)$, for $\nu_{\rm e}$ values in the meridial section at $12\,{\rm ms}$ in the laboratory frame ($rr$: top left; $\theta\theta$: top right; $\phi\phi$: middle left; $r\theta$: middle right; $r\phi$: bottom left; $\theta\phi$: bottom right). The neutrino energy $\epsilon$ is chosen to be the mean energy at each point. The plotted ranges are $0\,{\rm km}\,\le x\le 150\,{\rm km}$ and $-150\,{\rm km}\,\le z\le 150\,{\rm km}$. In each panel, the left and middle portions show the Boltzmann- and M1-Eddington tensors, respectively, whereas the right panels are the differences between the two, $k^{ij}_{\rm M1}(\epsilon)-k^{ij}(\epsilon)$. The off-diagonal components are multiplied by a factor of $10$ or $100$ as indicated at the bottom of each panel, in order to show them in similar color scales, which are different from panel to panel in fact.}
\end{figure*}

In figure \ref{fig:eddmap}, we compare the spatial distributions of the individual components between the Boltzmann- and M1-Eddington tensors for $\nu_{\rm e}$ with the mean energy in the laboratory frame at $12\,{\rm ms}$. The edge of the oval shape seen in each panel roughly corresponds to the shock surface (see the top middle panel of figure \ref{fig:snapshots}). All the diagonal components approach $1/3$ at the center for both the Boltzmann- and M1-Eddington tensors. This is consistent with the optically thick limit. The values of the $rr$-components then rise with radius to unity, whereas those of the $\theta\theta$- and $\phi\phi$-components decline to zero, which is again as expected in the optically thin limit. In between the transition from one limit to the other occurs in both cases, but it happens at a bit smaller radius for the M1-Eddington tensor as illustrated in the top two and middle left panels for the diagonal components.

\begin{figure}[tbph]
\centering
\includegraphics[width=\hsize]{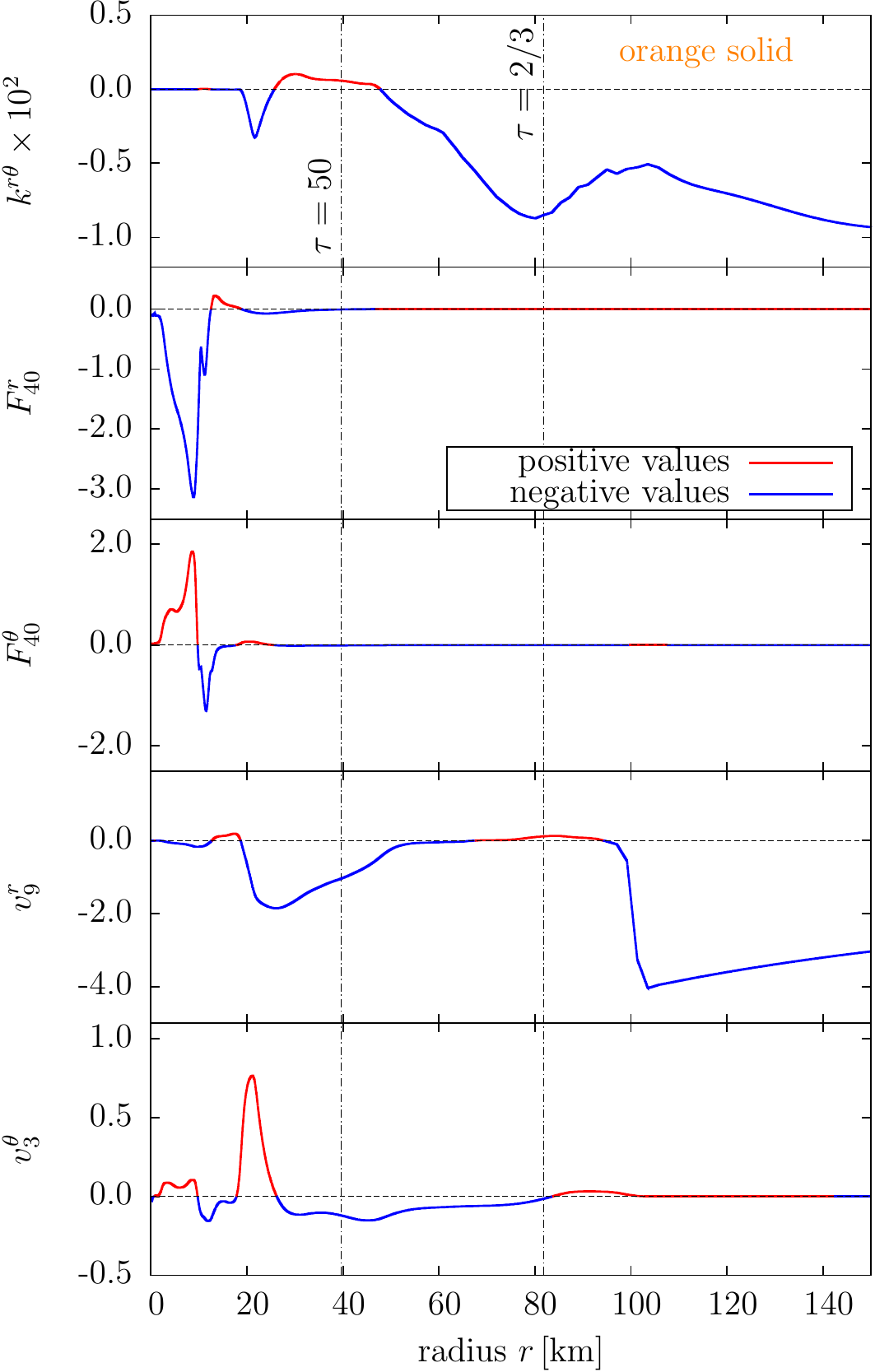}
\caption{\label{fig:ray115} Radial profiles of the $r\theta$-component of the Eddington tensor $k^{r\theta}$, the $r$- and $\theta$-components of the energy flux $F^i$ and the matter velocity $v^i$ along the orange solid line drawn in the middle right panel of figure \ref{fig:eddmap}. For the Eddington tensor and the flux, the neutrino energy is the mean energy at each point. The definitions of symbols are as follows: $F^r_{40}:=(F^r/10^{40}\,{\rm erg\,cm^{-2}\,s^{-1}})$, $F^\theta_{40}:=(F^\theta/10^{40}\,{\rm erg\,cm^{-2}\,s^{-1}})$, $v^r_{9}:=(v^r/10^{9}\,{\rm cm\,s^{-1}})$, and $v^\theta_{3}:=(v^\theta/10^{3}\,{\rm rad\,s^{-1}})$. In each panel, the portions of lines whose values are positive (negative) are colored red (blue) as indicated in the legend. The vertical dot-dashed lines correspond to the radii at which the optical depths for the average neutrino energy along the specified radial ray are $50$ and $2/3$ as indicated near the lines. }
\end{figure}

\begin{figure}[tbph]
\centering
\includegraphics[width=\hsize]{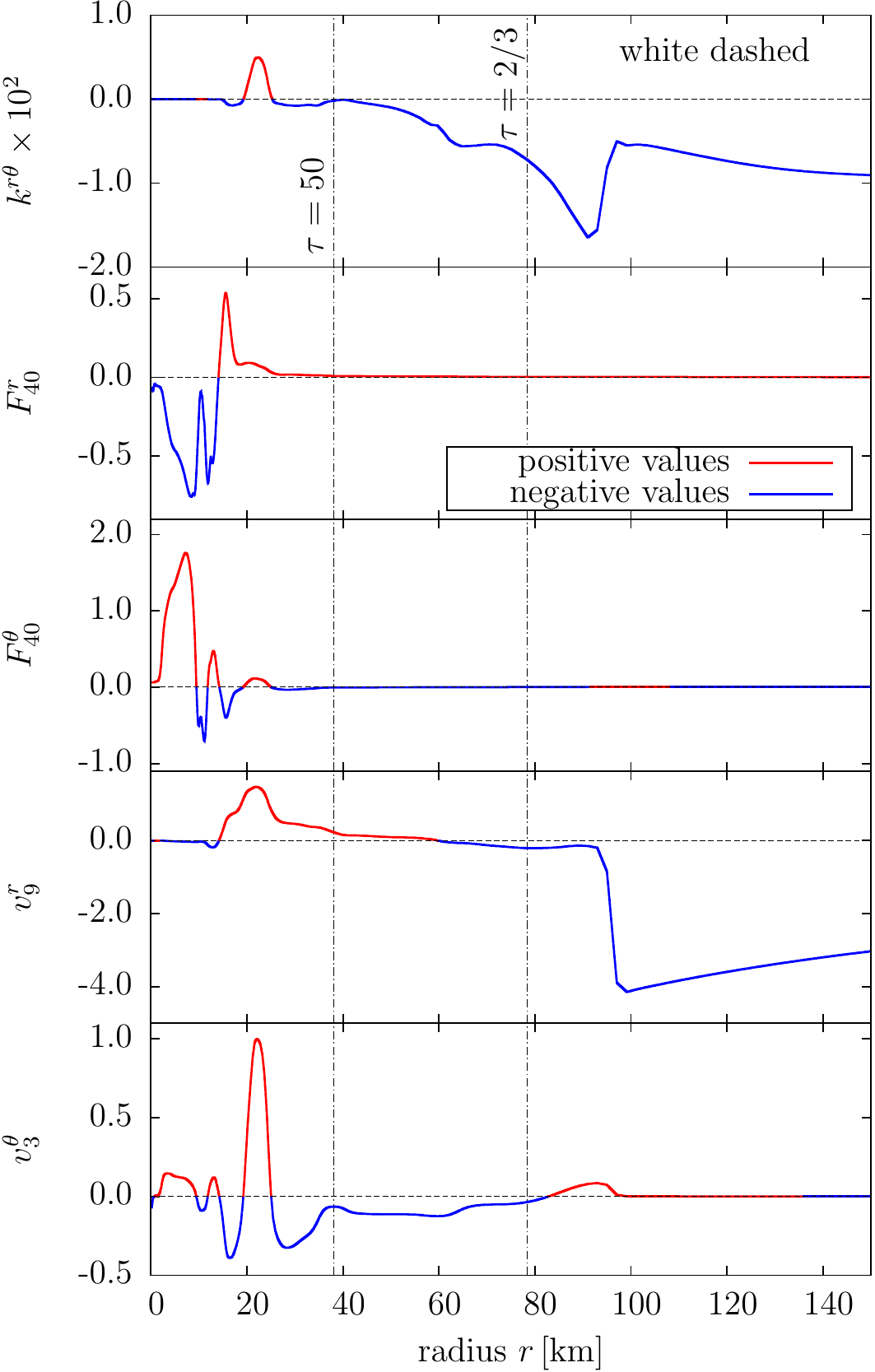}
\caption{\label{fig:ray96} Same as figure \ref{fig:ray115} but along with the white dashed line shown in figure \ref{fig:eddmap}.}
\end{figure}

\begin{figure}[tbph]
\centering
\includegraphics[width=\hsize]{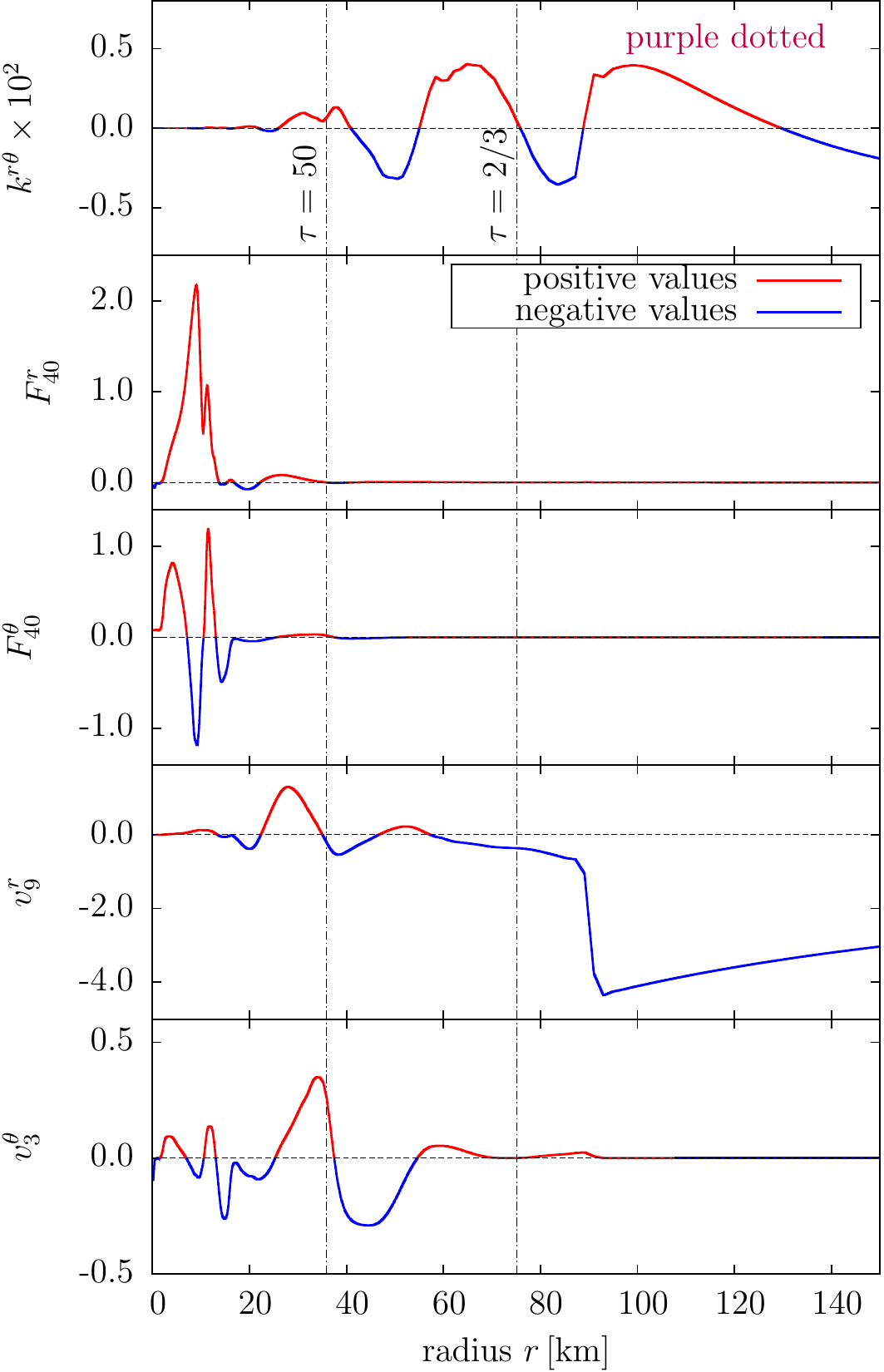}
\caption{\label{fig:ray79} Same as figures \ref{fig:ray115} and \ref{fig:ray96} but along with the purple dotted line given in figure \ref{fig:eddmap}.}
\end{figure}

Although the values of the off-diagonal components are not very large, being typically $\sim 1/10\text{--}1/100$ the diagonal components, their presence implies that the principal axes of this tensor do not coincide with the $r$-, $\theta$-, and $\phi$-directions. The behavior of the off-diagonal components is determined by complex combinations of matter motions and neutrino reactions. In order to show this, the profiles along some arbitrarily chosen radial rays are shown for several quantities of relevance in figures \ref{fig:ray115}--\ref{fig:ray79}. 

Since the Eddington tensor is the second angular moment of the distribution function, it is nothing but the amplitude of the $\ell=2$ mode in the spherical harmonics expansion, while the flux is the first angular moment and $\ell=1$ mode amplitude. Although the two modes are independent of each other in principle, they are correlated one way or another in reality. In the simplest case, for example, where a single bunch of neutrinos flies in one direction having, say, positive $r$- and $\theta$-components of flux, then the $r\theta$-component of the Eddington tensor should be positive. This is not true in general for multibunch cases, though. Keeping this simple fact in mind, we will look into the details of these figures.

In the optically thick region (optical depth, say, $\tau \gtrsim 50$), neutrinos are trapped by matter and they move in tandem. The relativistic aberration tilts the neutrino distribution so that the neutrino flux should be aligned with the matter velocity. From the inspection of the second to fifth panels of figures \ref{fig:ray115}--\ref{fig:ray79}, one finds that the signs of the $r$- and $\theta$-components of the neutrino flux coincide with those of the matter velocity counterparts. The sign of $k^{r\theta}$ is identical to that of the product of $v^rv^\theta$ or $F^rF^\theta$, since neutrinos are comoving with matter in unison in the optically thick region.

In the semitransparent (optical depth $2/3\lesssim \tau \lesssim 50$) region, the sign of $k^{r\theta}$ still coincides with that of the product $F^r F^\theta$, again indicating that the Eddington tensor is correlated with the flux. On the other hand, the $r$-components of the neutrino flux and the matter velocity have opposite signs, whereas their $\theta$-components have the same sign. This is because interactions between neutrinos and matter are no longer strong enough to enforce the comoving of neutrinos with matter in the radial direction.

\begin{figure}[tbph]
\centering
\includegraphics[width=\hsize]{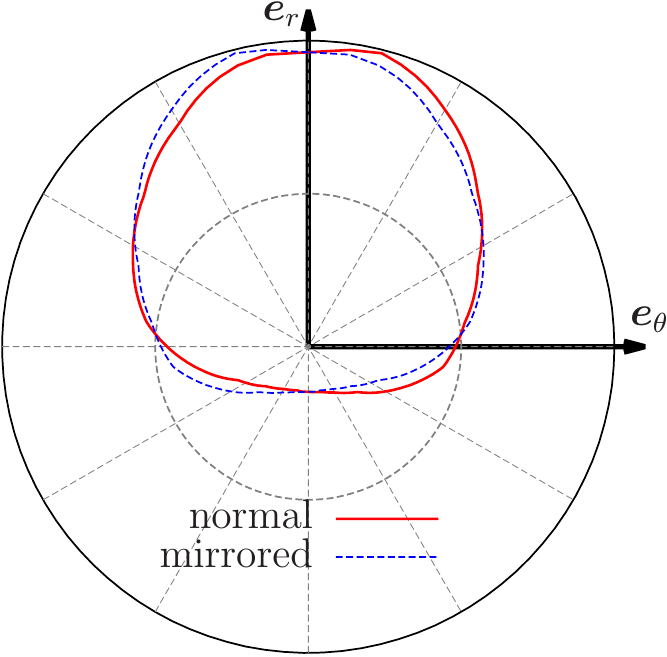}
\caption{\label{fig:sectionwire} Angular distribution of $\nu_{\rm e}$ on the plane spanned by $\bs e_r$ and $\bs e_\theta$ in momentum space at $r=82\,{\rm km}$ on the radial ray given as the purple dotted line in figure \ref{fig:eddmap}. The neutrino energy is set to the mean energy ($\sim 11\,{\rm MeV}$) at this point. Note that the energy and angle are measured in the laboratory frame. The red solid and blue dashed curves are the original distribution and its mirror image with respect to the $\bs e_r$-axis, respectively.}
\end{figure}

In the optically thin (optical depth $\tau \lesssim 2/3$) region, the correlation between the flux and the Eddington tensor is not simple. In fact, there are regions along the three radial rays in figures \ref{fig:ray115}--\ref{fig:ray79}, where both the $r$- and $\theta$-components of the flux are positive while the $k^{r\theta}$ is negative. This implies that there are multiple bunches of neutrinos that are moving differently, which can be understood by looking at the distribution function. Shown in figure \ref{fig:sectionwire} is not only the angular distribution of $\nu_{\rm e}$ at the point in the optically thin region along the purple line in figure \ref{fig:eddmap} but also its mirror image in order to emphasize the nonaxisymmetric distortion.

It is evident from the figure that the neutrinos are mainly flying in upper right direction. It should also be apparent that there are some neutrinos moving in the lower right direction. The former component is neutrinos coming from the PNS, bent by matter in the semitransparent regions, whereas the latter component is emitted from the neighborhood. They are beamed by the matter motion. As a matter of fact, the matter velocity is $v^r < 0$ and $v^\theta > 0$ at $r=82\,{\rm km}$ (see figure \ref{fig:ray79}), the same direction as the latter component. Hereafter the former is called the PNS component and the latter is called the neighborhood component.

As for the corresponding component of the Eddington tensor $k^{r\theta}$, the neighborhood component is dominant over the PNS component along the purple dotted line in figure \ref{fig:eddmap}. As a result, its sign changes from that in the semitransparent region and returns to it again outside the shock. Along the orange line, on the other hand,  $v^r > 0$ and $v^\theta >0$ in the optically thin region (see figure \ref{fig:ray115}), which implies that the neighborhood component gives a positive contribution to $k^{r\theta}$. The fact that the actual value of $k^{r\theta}$ is negative indicates that the PNS component dominates it.
There is yet another case along the white line, in which $k^{r\theta}$ is negative while both $F^r$ and $F^\theta$ are positive in the optically thin region (see figure \ref{fig:ray96}). This happens because the PNS component gives a large positive contribution to $F^r$ and a small negative contribution to $F^\theta$, while the neighborhood component contributes in the opposite sense to $F^r$ and $F^\theta$. As a result, $k^{r\theta} < 0$, $F^r > 0$, and $F^\theta > 0$ are realized simultaneously.

\subsubsection{Comparison between Boltzmann- and M1-Eddington Tensors}

\begin{figure}[tbph]
\centering
\includegraphics[width=\hsize]{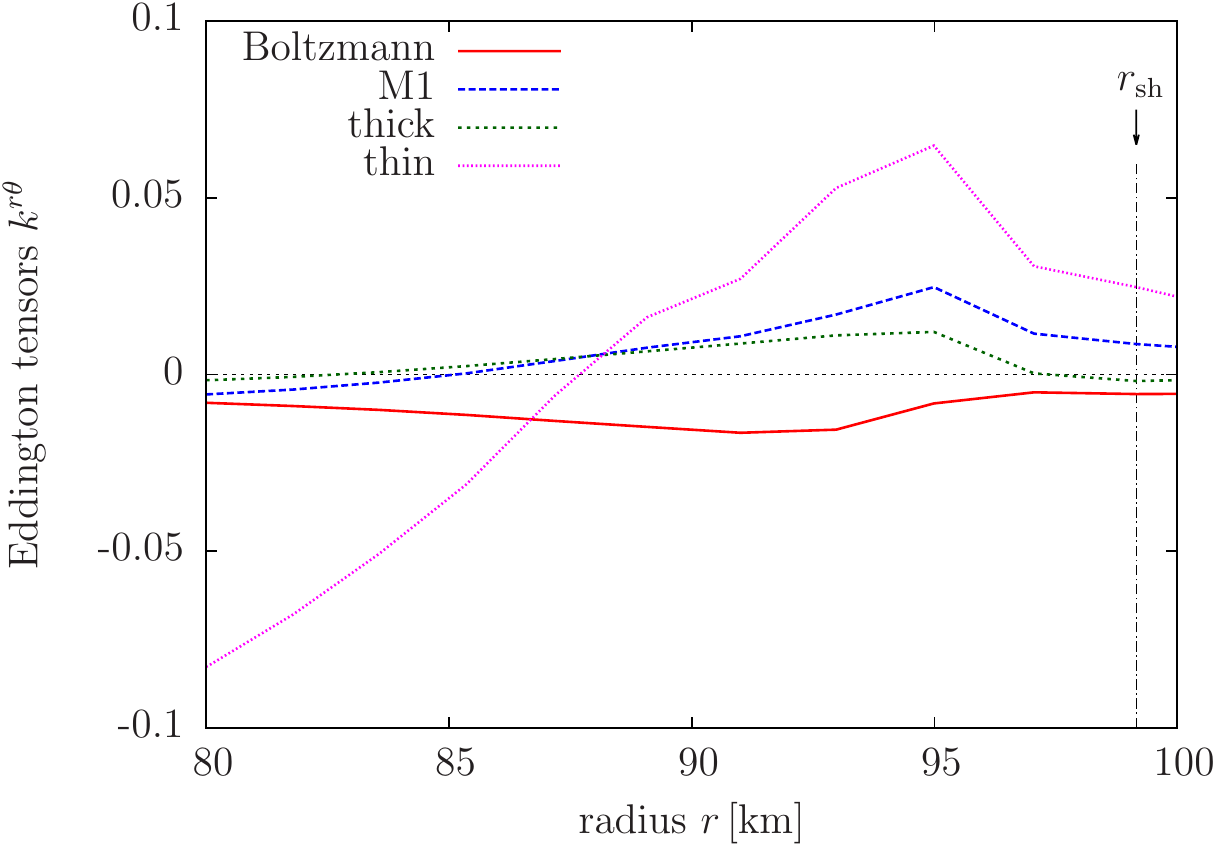}
\caption{\label{fig:BolMthinthick}  Radial profiles of the Eddington tensors $k^{r\theta}(\epsilon)$, $k^{r\theta}_{\rm M1}(\epsilon)$, $k^{r\theta}_{\rm thin}(\epsilon)$, and $k^{r\theta}_{\rm thick}(\epsilon)$ along $\theta = \pi/4$ (white dashed line in figure \ref{fig:eddmap}). The neutrino energy is the mean energy at each point. The vertical dot-dashed line indicates the position of the shock.}
\end{figure}

Now we shift our attention to the comparison of the Boltzmann- and M1-Eddington tensors. Their off-diagonal components are very similar in the optically thick and thin limits. This is as expected because neutrinos are moving in unison in these cases (dragged by matter in the former and free streaming in the latter). Their behaviors are different in the semitransparent regions, however. As a matter of fact, the $r\theta$-components are different even in the signature near the shock whereas the values of the $r\phi$- and $\theta\phi$-components for the M1-Eddington tensor are twice as large as those for the Boltzmann-Eddington tensor in the same region.

We show in figure \ref{fig:BolMthinthick} radial profiles of the $r\theta$-components for the Boltzmann-Eddington tensor $k^{r\theta}$ (equation (\ref{eq:boltzpij})) and for the M1-Eddington tensor $k^{r\theta}_{\rm M1}$ (equation (\ref{eq:m1pij})) together with the optically thin limit $k^{r\theta}_{\rm thin}:=P^{r\theta}_{\rm thin}/E$ (equation (\ref{eq:thinpij})) and optically thick limit $k^{r\theta}_{\rm thick}:=P^{r\theta}_{\rm thick}/E$ (equation (\ref{eq:thickpij})) used in the prescription of the M1-Eddington tensor. In the figure, $k^{r\theta}$ is always negative, whereas $k_{\rm M1}^{r\theta}$, $k_{\rm thick}^{r\theta}$, and $k_{\rm thin}^{r\theta}$ are not. One finds that both $k^{r\theta}_{\rm M1}$ and $k^{r\theta}_{\rm thick}$ become positive at $r\gtrsim 85\,{\rm km}$ while $k^{r\theta}_{\rm thin}$ gets positive slightly farther out at $r\sim 87\,{\rm km}$. As indicated in figure \ref{fig:ray96}, $F^r$ is consistently positive in these regions, whereas $F^\theta$ changes sign from positive to negative at $r\sim87\,{\rm km}$. As a consequence, the optically thin limit of the M1-Eddington tensor mistakenly takes positive values at $r\gtrsim 87\,{\rm km}$. On the other hand, $k^{r\theta}_{\rm thick}$ takes positive values inside this radius. This is because the sum $H^r V^\theta + V^r H^\theta$ in equation (\ref{eq:thickpij}) is positive. It is worth noting that in equation (\ref{eq:thickpij}) some correction terms whose order with respect to the local mean free path is higher than zeroth are neglected. The wrong sign of $k_{\rm thick}^{r\theta}$ indicates that higher-order corrections cannot be neglected in this region. The M1-closure method tries to correct such errors by interpolating the optically thick and thin limits, however. The results shown in the figure demonstrate that the attempt fails here. The errors in the off-diagonal components of the Eddington tensor may affect the lateral component of the neutrino flux as discussed in \cite{2018ApJ...854..136N}.

\begin{figure*}[tbph]
\centering
\includegraphics[width=\hsize]{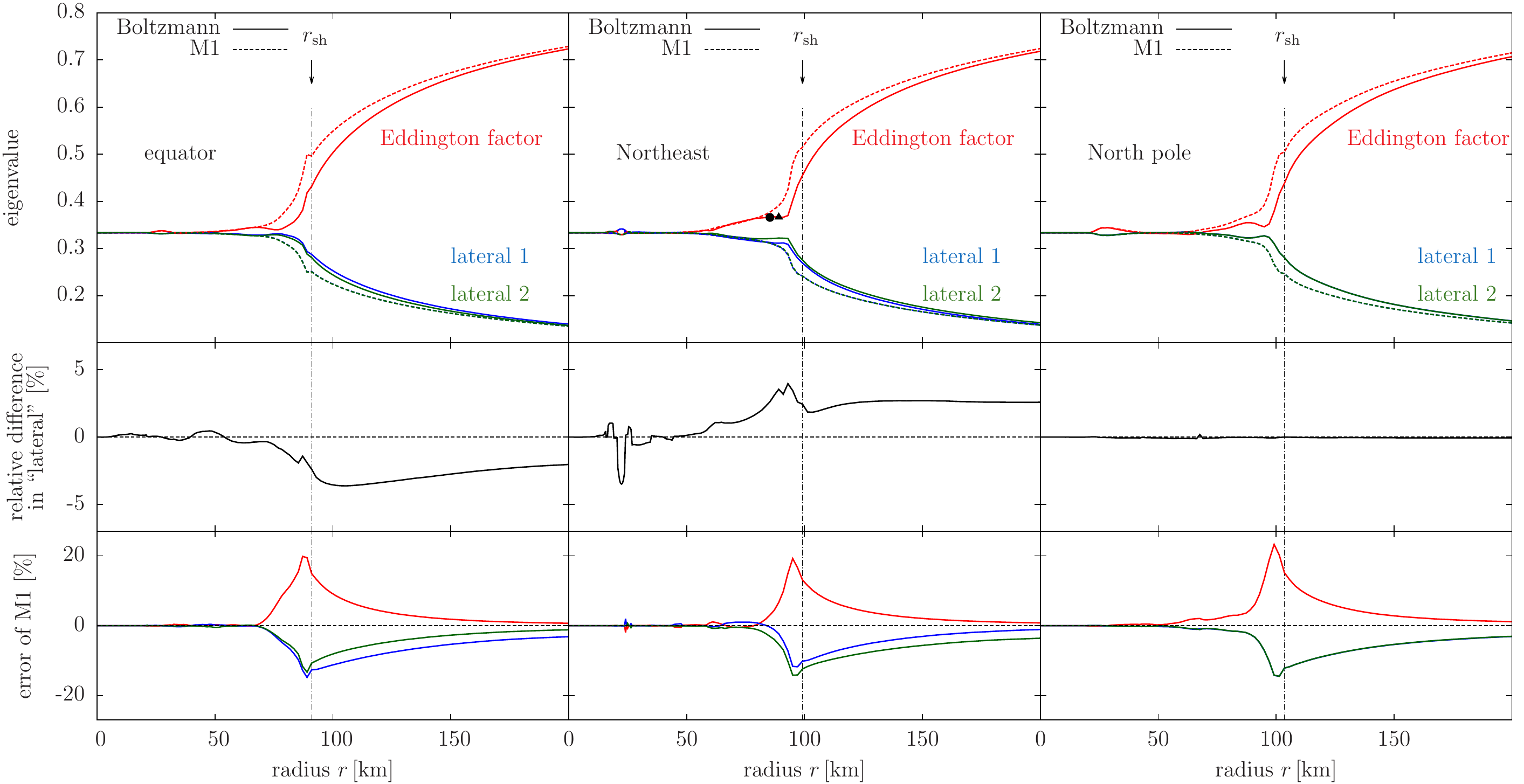}
\caption{\label{fig:eigenvalue} Radial profiles of eigenvalues for the Boltzmann- and M1-Eddington tensors at $12\,{\rm ms}$ after bounce in the laboratory frame for the electron-type neutrinos with their mean energies. The largest eigenvalue, or the Eddington factor, is shown in red, and the other two eigenvalues named ``lateral 1, 2'' are represented in blue and green, respectively. The solid and dashed lines correspond to the eigenvalues of the Boltzmann- and M1-Eddington tensors, respectively. The directions of the radial rays chosen here are $\theta=\pi/2$ (equator; left panels), $\theta = \pi/4$ (northeast; middle panels), and $\theta = 0$ (north pole; right panels). In each panel, the shock radius for the particular direction is indicated with the vertical dot-dashed line. In the top panels the eigenvalues themselves are presented, whereas the relative differences between ``lateral 1, 2'' of the Boltzmann-Eddington tensor are shown in the middle panels, and the fractional differences between the Boltzmann- and M1-Eddington tensors are displayed in the bottom panels. The black small circle and triangle in the top middle panel are the points where the Boltzmann-Eddington factors are similar but the M1-Eddington factors are different. The reason for this behavior is explained by comparing the angular distributions in figure \ref{fig:Ed_dis}.}
\end{figure*}

Since the Eddington tensor is a symmetric tensor, it can always be diagonalized and its eigenvalues and eigenvectors characterize the shape of the distribution function. The largest eigenvalue, or the Eddington factor, represents how sharp the distribution is along the principal direction, and the other two eigenvalues indicate how flat it is in the perpendicular directions. We hence show the eigenvalues of the Boltzmann- and M1-Eddington tensors in figure \ref{fig:eigenvalue}. One finds again that the Eddington factor takes the optically thick limit of $1/3$ deep inside the core and increases toward the shock, and it reaches the free-streaming limit outside it. Since the sum of three eigenvalues of the Eddington tensor should be unity (see equation (14) in \citet{1984JQSRT..31..149L}), two other eigenvalues, which are positive normally, decrease with radius.

As stated above, the M1-closure method assumes the axisymmetric distribution with respect to the flux direction. As a result, two eigenvalues other than the Eddington factor in the M1-Eddington tensor are degenerate (blue and green dashed lines denoted by ``lateral 1'' and ``lateral 2'') in figure \ref{fig:eigenvalue}. These lateral eigenvalues of the Boltzmann-Eddington tensor, on the other hand, are slightly different from each other, since no symmetry is imposed artificially on the neutrino distribution in our simulations. However, the difference between lateral 1 and 2, which is defined as $(\kappa_{\rm lat2} - \kappa_{\rm lat1})/\kappa_{\rm lat1}$ with $\kappa_{\rm lat1,2}$ being the eigenvalues of lateral 1, 2, is only a few percent typically as shown in the middle panels of figure \ref{fig:eigenvalue}, indicating that the axisymmetry with respect to the flux direction is nearly achieved as a consequence of the evolution.

The estimation of the Eddington factor in the M1-closure method is not so accurate. The fractional differences between the Boltzmann- and M1-Eddington tensors, which are defined as $(\kappa_{\rm M}-\kappa_{\rm B})/\kappa_{\rm B}$ for their corresponding eigenvalues $\kappa_{\rm B}$ and $\kappa_{\rm M1}$, are also presented in the bottom panels of figure \ref{fig:eigenvalue}. It is found that the fractional difference reaches $\sim 20\%$, just behind the shock. Note that although there are some alternatives to the approximate functions in equation (\ref{eq:levermore}) \citep[e.g.,][]{2015MNRAS.453.3386J}, we still find $\sim 10\%$ of maximum errors at least for them.

In the vicinity of the black small circle and triangle in the top middle panel of figure \ref{fig:eigenvalue}, the M1-Eddington factor increases although the Boltzmann counterpart stays at almost the same value or even decreases slightly with radius just behind the shock. Since the M1-Eddington factor given in equation (\ref{eq:levermore}) is a monotonically increasing function of the flux factor, the latter also increases when the Eddington factor does not. The key to the understanding of such behaviors is the distribution function again.

\begin{figure}[tbph]
\centering
\includegraphics[width=0.8\hsize]{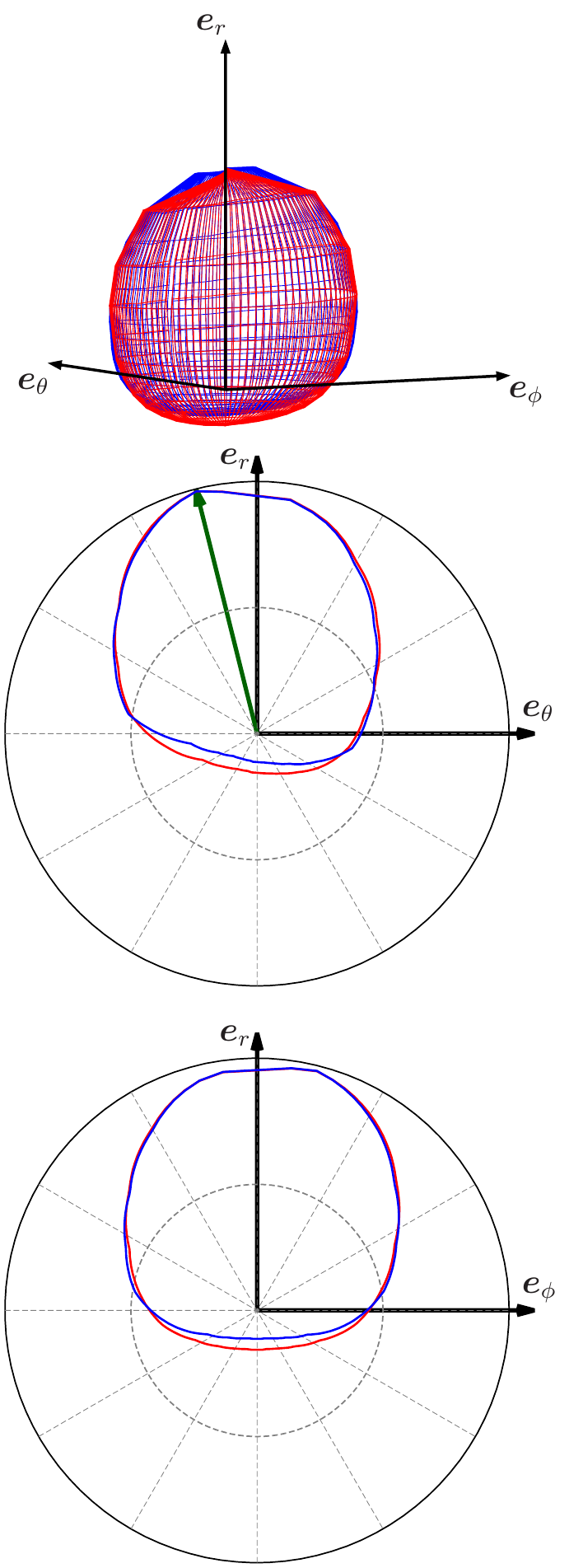}
\caption{\label{fig:Ed_dis} Angular distributions in momentum space of electron-type neutrinos at the two points in the vicinity of the shock that are marked with the small circle and triangle in the top middle panel of figure \ref{fig:eigenvalue}. The neutrino energies are again set to the mean energies at the individual points. The energy and angle of neutrinos are measured in the laboratory frame. Each distribution is normalized with its maximum value. The red and blue colors represent the quantities at the circle and triangle points, respectively. The top panel shows the three-dimensional angular distribution, and the middle and bottom panels display those on the sections spanned by $\bs e_r$--$\bs e_\theta$ and $\bs e_r$--$\bs e_\phi$, respectively. The green arrow in the middle panel is the flux direction defined in the text.}
\end{figure}

Shown in figure \ref{fig:Ed_dis} are two angular distributions of electron-type neutrinos, which are taken at the positions of the small black circle and triangle put in the top middle panel of figure \ref{fig:eigenvalue}. In the following discussions, we refer to the ``flux direction'' as the direction in which the distribution function is the maximum \footnote{This ``flux direction'' might not coincide with the direction of the flux $F^i$, since the latter is determined not by the maximum value but by the angular average of the unit vector.}. It is shown as the green arrow in the middle panel of figure \ref{fig:Ed_dis}. It is found that the distribution function opposite to the flux direction is a bit smaller at the point of the triangle, which is closer to the shock. Since, roughly speaking, the flux factor and the Eddington factor are proportional to $\langle \cos \tilde{\theta} \rangle$ and $\langle \cos^2 \tilde{\theta} \rangle$, respectively, where $\tilde{\theta}$ is the zenith angle with respect to the flux direction and $\langle \cdot \rangle$ represents the average over the solid angle, the reduction of the distribution on the opposite side of the flux direction, $\cos \tilde{\theta} \sim -1$, leads to the larger flux factor and slightly smaller Eddington factor at the triangle position than at the circle position. In fact, there is a subtlety here. Since the solid-angle average is given as $\langle \cdot \rangle := \int f \cdot \rd \Omega_p / \int f \rd \Omega_p$, the reduction of $f$ at $\cos \tilde{\theta} \sim -1$ always results in a decrease of the denominator and hence necessarily leads to an increase in the flux factor, whereas the Eddington factor is not much changed.

What is important here is the fact that only the backward portion ($\cos \tilde{\theta} \sim -1$) in the normalized angular distributions is depleted. This situation is induced by the emissions from the neighborhood. 
Note that our Boltzmann code can treat this situation properly, since the forward- and backward-propagating neutrinos are treated individually. This is not the case for the M1-closure method, though, since it treats only the angle-averaged quantities and does not distinguish the increase in forward-propagating neutrinos from the decrease in backward-propagating neutrinos. If additional information on the emission from the neighborhood is somehow incorporated in the approximate formula of equation (\ref{eq:levermore}), the M1-closure method may be improved. That is beyond the scope of this paper, however.

\subsubsection{Resolution}

\begin{figure*}[tbph]
\centering
\includegraphics[width=\hsize]{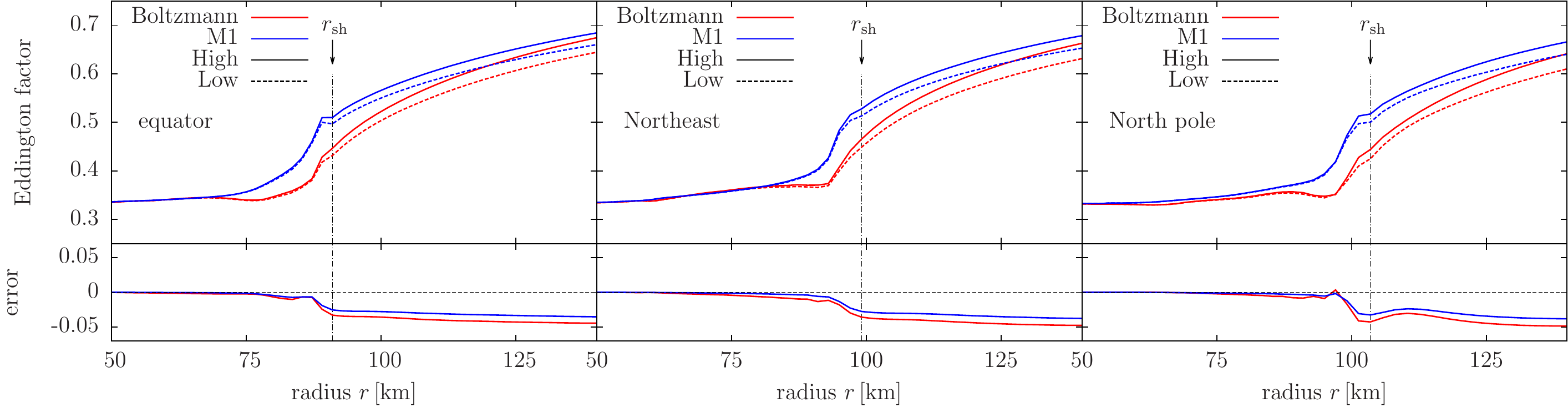}
\caption{\label{fig:reschk} Comparison of the Boltzmann-Eddington (red) and M1-Eddington (blue) factors for the simulations with the higher (solid lines) and lower (dotted lines) resolutions. Only the electron-type neutrinos are shown at their mean energies. In the lower panels, the fractional differences between the higher- and lower-resolution results are plotted.}
\end{figure*}

Due to the limited computational resources, the angular resolution in momentum space is not very high in our simulations. We refer readers to \citet{2017ApJ...847..133R} for detailed discussions on the issues of resolution and convergence. In the optically thick regime where the neutrino distribution is almost isotropic, this limited resolution does not pose a serious problem since such distributions can be accurately expressed with a small number of angular bins. On the other hand, in the optically thin regime, forward-peaked distributions cannot be correctly reproduced with poor resolutions. One may hence think justifiably that the differences shown in figure \ref{fig:eigenvalue} are mostly artifacts of the insufficient resolutions.

In order to check the resolution dependence, we run additional simulations with both lower and higher resolutions. In these simulations, we take and fix the matter distribution at $12\,{\rm ms}$ after bounce and compute only the neutrino distribution functions in steady states. In order to minimize the computational cost, we limit the computational domain to $\sim 40\,{\rm km} < r < \sim 300\,{\rm km}$. The numbers of the angular grid points in momentum space are $(\theta_\nu,\,\phi_\nu) = (10,\, 6)$ and $(14,\, 10)$ for the lower and higher resolutions, respectively.

Figure \ref{fig:reschk} shows the results of the additional simulations. It is similar to figure \ref{fig:eigenvalue}, but we plot only the Eddington factors and their fractional differences not between the Boltzmann- and M1-Eddington tensors but between the different-resolution calculations with the Boltzmann solver. It is clear that the fractional differences in both the Boltzmann- and M1-Eddington tensors are small in the optically thick region, especially where the Boltzmann-Eddington factor is $\le 0.4$. On the other hand, they are as large as $\sim 5\%$ in the semitransparent to optically thin regions. The numerical convergence is hence not yet reached in the outer regions. Note that this is consistent with the results in \citet{2017ApJ...847..133R}. What is more important here, however, is that the large difference observed between the Boltzmann- and M1-Eddington tensors in figure \ref{fig:eigenvalue} still exists in figure \ref{fig:reschk} (see the difference between red and blue lines). It is concluded, therefore, that this is not an artifact of the relatively low resolution in the Boltzmann simulations.

\subsection{Angular Momentum Transport}

We finally discuss the angular momentum that is carried away by neutrinos (see figure \ref{fig:angularmomentum}). 
It is evaluated from the distribution function directly. The energy--momentum tensor of neutrinos is defined as
\begin{equation}
T^{\sigma\rho}_{(\nu)} : = \int f p^\sigma p^\rho \rd V_p, \label{eq:emtennu}
\end{equation}
and satisfies the conservation law,
\begin{equation}
\nabla_\sigma T^{\sigma\rho}_{(\nu)} = G^\rho,
\end{equation}
where $G^\rho$ is defined in equation (\ref{eq:interaction}). Note that the energy--momentum tensor is also expressed as $T^{\sigma\rho}_{(\nu)} = \int M^{\sigma\rho}(\epsilon) \rd (\epsilon^3/3)$. Using the Killing vector $\xi = \pa_\phi$ that exists under axisymmetry, we can define the angular momentum $4$-current as $j^\rho := \xi_\sigma T^{\sigma \rho}$, which obeys the angular momentum conservation, 
\begin{equation}
\nabla_\rho j^\rho = \xi_\rho G^\rho.
\end{equation}
Defining the angular momentum of neutrinos inside the sphere of radius $r$ as $J_{(\nu)}(r) := \int_0^r j^t \rd V_x$, we write the conservation law in the integral form,
\begin{equation}
\dot{J}_{(\nu)}(r) + \int_{S(r)} j^r \rd s = \int_0^r \xi_\rho G^\rho \rd V_x,
\end{equation}
where $\rd s$ is the surface element.
The right-hand side represents the exchange of angular momentum between neutrinos and matter. Assuming that advection of the angular momentum of matter is negligible, then we can evaluate the angular momentum loss by neutrinos from the sphere as
\begin{equation}
\dot{J}(r) := -\int_{S(r)} j^r \rd s = -\int_{S(r)} r^2 \sin^2\theta T^{\phi r}_{(\nu)} \rd s. \label{eq:angmomdistri}
\end{equation}
In the discussions below, we set $r=100\,{\rm km}$ since the numerical resolution poses no problem up to this radius (see figure \ref{fig:reschk}). Not to mention, we take a sum over all neutrino flavors.

\citet{1978ApJ...219L..39E} proposed a way to analytically estimate the angular momentum loss by neutrinos. It is expressed in the natural unit as
\begin{equation}
\dot{J} = -\int \left(\frac{L_\nu}{4\pi r^2}\right)\omega r_\perp^2 \rd s, \label{eq:angmomepstein}
\end{equation}
where $L_\nu$ and $r_\perp$ are the neutrino luminosity and the length of the lever from the rotational axis, respectively, and the integral is done over the ``stellar surface" where neutrinos are emitted. In the current context, it should be interpreted as the neutrinosphere. In the derivation, he assumes that the neutrino distributions are isotropic in the fluid rest frame and acquires anisotropy in the laboratory frame solely from the relativistic beaming by the rotation of matter. In equation (\ref{eq:angmomepstein}), we also need the neutrino luminosity. We adopt the blackbody formula for each neutrino flavor at the neutrinosphere, 
\begin{equation}
L_\nu = 4\pi r^2 \times \frac{7}{16} \sigma_{\rm SB} T^4, \label{eq:fermilumi}
\end{equation}
since the formula was originally meant to be used that way. In this expression, $\sigma_{\rm SB}$ and $T$ are the Stefan--Boltzmann constant and the matter temperature at the neutrinosphere, respectively. In the following evaluation, the neutrinosphere is set at the radius where the density is $\rho = 10^{11}\,{\rm g\,cm^{-3}}$. In other words, the surface integral is conducted over the isodensity surface with $\rho = 10^{11}\,{\rm g\,cm^{-3}}$ and multiplied by six to account for the six neutrino flavors: $\nu_{\rm e}$, $\bar{\nu_{\rm e}}$, $\nu_\mu$, $\bar{\nu_\mu}$, $\nu_\tau$, and $\bar{\nu_\tau}$.

\begin{figure}[tbph]
\centering
\includegraphics[width=\hsize]{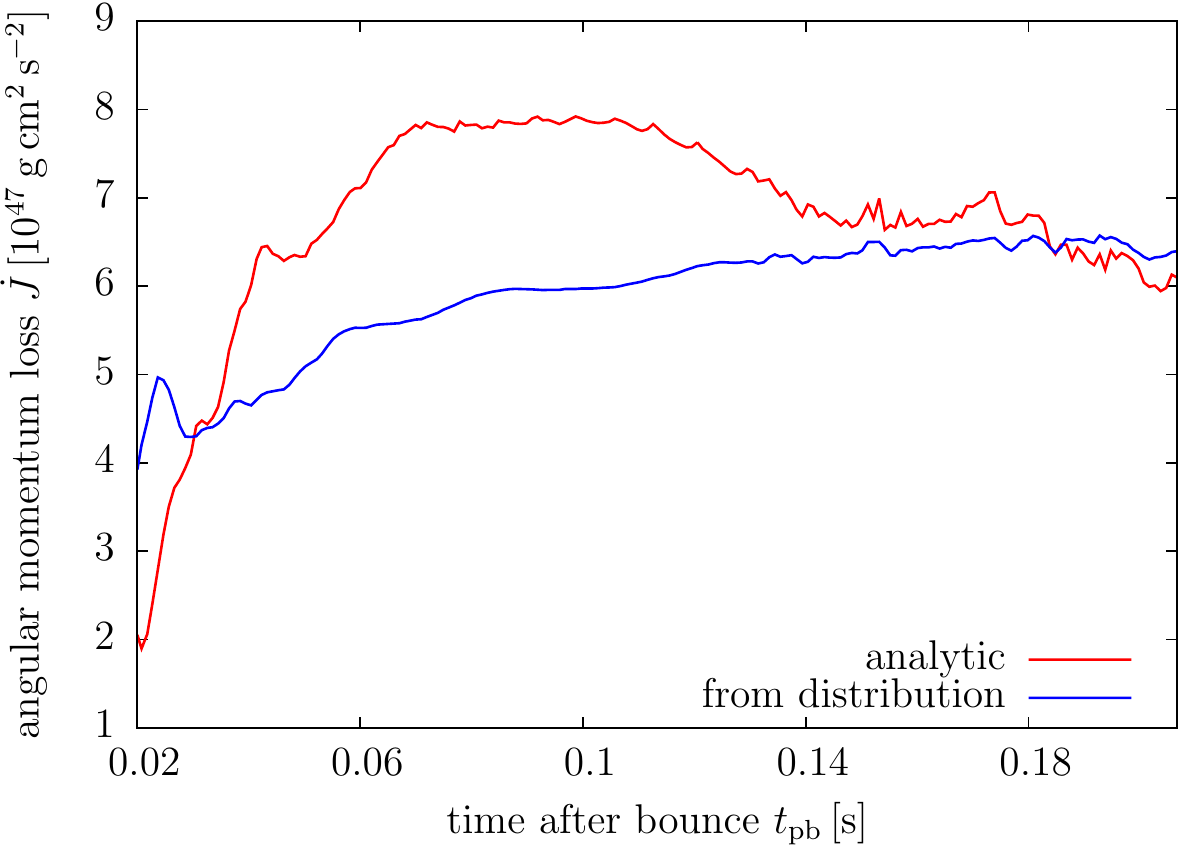}
\caption{\label{fig:angmomloss} Angular momentum loss by neutrino emissions as a function of the time after bounce. The blue and red lines show the evaluation from equations (\ref{eq:angmomdistri}) and (\ref{eq:angmomepstein}), respectively.}
\end{figure}

In figure \ref{fig:angmomloss}, we compare the angular momentum losses estimated from equations (\ref{eq:angmomdistri}) and (\ref{eq:angmomepstein}). Since equation (\ref{eq:angmomdistri}) is evaluated at $r=100\,{\rm km}$, we plot the evolutions only after the time when the minimum shock radius exceeds that radius. It is found that the evaluation of equation (\ref{eq:angmomdistri}) gives a much more gradual increase than the estimate from equation (\ref{eq:angmomepstein}) and the deviation reaches $\sim 30\%$ around $100\,{\rm ms}$ after bounce. Although this is not small, the analytical formula is good enough to obtain the order of magnitude of the angular momentum loss, indicating that the basic picture of the angular momentum loss via neutrino emission is correctly described by \citet{1978ApJ...219L..39E}.

\section{Summary and Discussion} \label{sec:summary}
In order to examine the effects of rotation on the supernova dynamics and, in particular, on the distributions of neutrinos, we performed a core-collapse simulation for a rotating progenitor with the Boltzmann-radiation-hydrodynamics code. 
Although the shock morphology is different, the average shock radius, the luminosities, and the mean energies of neutrinos for the modest rotation we assumed in this paper are not much different from those in the corresponding nonrotating model presented in \citet{2018ApJ...854..136N}. Besides, no successful shock revival is obtained. This result is consistent with \cite{2018ApJ...852...28S}.

The neutrino distributions are affected by the rotation, though. The relativistic aberration tilts the neutrino distributions in the rotational direction. As a consequence, the azimuthal component of the neutrino flux emerges. It is interesting that this component is positive, i.e., has the same sign as $v^\phi$, in the laboratory frame, whereas it is negative in the fluid rest frame, meaning that matter is rotating faster than neutrinos.

Then we compared the Eddington tensor obtained directly from our Boltzmann simulation with that evaluated according to the M1-closure prescription from the same data. The Eddington tensor is determined by some complicated combinations of the matter velocity, local neutrino reactions, and the neutrino flux that originated deeper inside. We found the earlier transition from the optically thick to thin limits for the diagonal components of the M1-Eddington tensor. The behavior of the off-diagonal components is quantitatively (for the $r\phi$- and $\theta\phi$-components) and even qualitatively (for the $r\theta$-component) different in the semitransparent region. The deviation in the Eddington factors reaches $\sim 20\%$ just behind the shock. The discrepancy is originated from the poor performance of the M1-closure prescription for the particular angular distributions of neutrinos in momentum space, in which only the neutrinos going almost in the opposite direction to the flux direction are depleted. We found in such cases that the flux factor is increased but the Eddington factor is decreased and, as a result, the M1-Eddington factor increases while the Boltzmann-Eddington factor decreases. In order to correct such a qualitatively wrong behavior in the M1-closure prescription, we have to somehow take into account the effect of emissions from the neighborhood better. Although the resolution in our Boltzmann simulation is rather low, the discrepancy in the Eddington tensors is not an artifact of the resolution since it is also found in the high-resolution simulation.

Finally, the angular momentum loss by neutrino emissions was evaluated both directly from the distribution functions and analytically according to the Epstein formula. It is found that the latter approximation tends to overestimate the angular momentum loss but that the error is at the level of several tens of percent.

In this paper we discussed effects of rotation, assuming axisymmetry. New features may appear in 3D simulations. \citet{2016MNRAS.461L.112T}, for example, reported that the nonaxisymmetric fluid instability called low-$T/|W|$ instability revives the stalled shock in their 3D models. Such an instability may also occur in 3D simulations with the Boltzmann solver, changing the dynamics significantly. The 3D version of our Boltzmann-radiation-hydrodynamics code is under development, and results of such an investigation will be reported elsewhere. 
Although we studied only a modestly rotationg model in this paper, faster rotations are certainly our concern. Then not only the neutrino distributions but the dynamics itself will also be affected. For instance, the rotational core bounce, which is induced not by nuclear forces but by centrifugal forces, is an interesting topic. We are currently running such simulations at present, and the results will be published later.

The improvement of our code is also underway. Among other things, how the general relativistic strong gravity affects the supernova dynamics, as well as the distributions of neutrinos, is our concern. Note that the Boltzmann solver described in \cite{2017ApJS..229...42N} and used in this paper has already implemented general relativity in the $3+1$ decomposition of spacetime, although only the uniform acceleration of the entire system in the flat spacetime has been employed. Some tests in curved spacetimes and/or the coupling with dynamical spacetimes will be reported in a forthcoming paper. We have so far developed a numerical relativity module in polar coordinates like what is proposed in \citet{2013PhRvD..87d4026B}, which will also be published elsewhere soon.

\acknowledgments
We thank Yu Yamamoto for providing us with a subroutine that calculates internal energy of electrons and positrons. 
This research used high-performance computing resources of the K-computer and the FX10 of the HPCI system provided by the AICS and the University of Tokyo through the HPCI System Research Project (Project ID: hp160071, hp170031, and hp180111), the Computing Research Center in KEK, JLDG on SINET4 of NII, the Research Center for Nuclear Physics at Osaka University, the Yukawa Institute of Theoretical Physics at Kyoto University, and the Information Technology Center at University of Tokyo.
This work was supported by the Grant-in-Aid for Scientific Research (26104006, 15K05093, 16H03986), Grant-in-Aid for Innovative Areas (24103006), and Grant-in-Aid for Scientific Research on Innovative areas "Gravitational wave physics and astronomy:Genesis" (17H06357, 17H06365) from the Ministry of Education, Culture, Sports, Science and Technology (MEXT), Japan.
This work was also partly supported by research programs at K-computer of the RIKEN AICS, HPCI Strategic Program of Japanese MEXT, “Priority Issue on Post-K-computer” (Elucidation of the Fundamental Laws and Evolution of the Universe), and Joint Institute for Computational Fundamental Sciences (JICFus).
A.H. was supported in part by the Advanced Leading Graduate Course for Photon Science (ALPS) at University of Tokyo and Grant-in-Aid for JSPS Research Fellow (JP17J04422). H.N. was supported partially by JSPS Postdoctoral Fellowships for Research Abroad No. 27-348, Caltech through NSF award No. TCAN AST-1333520, and DOE SciDAC4 Grant DE-SC0018297 (subaward 00009650). 

\software{gnuplot \cite{gnuplot}}

\appendix
\section{Some Diagnostics} \label{sec:diagnostics}
Although the rotation in our simulation is modest and does not essentially affect the dynamics, some diagnostics are still useful for the comparison with other works. In this appendix, we hence present the ratio of the rotational energy to the gravitational energy, the electron fraction as a function of the density, the timescale ratio, and the trajectory of the PNS center, for that purpose.

\begin{figure}[tbph]
\centering
\includegraphics[width=0.5\hsize]{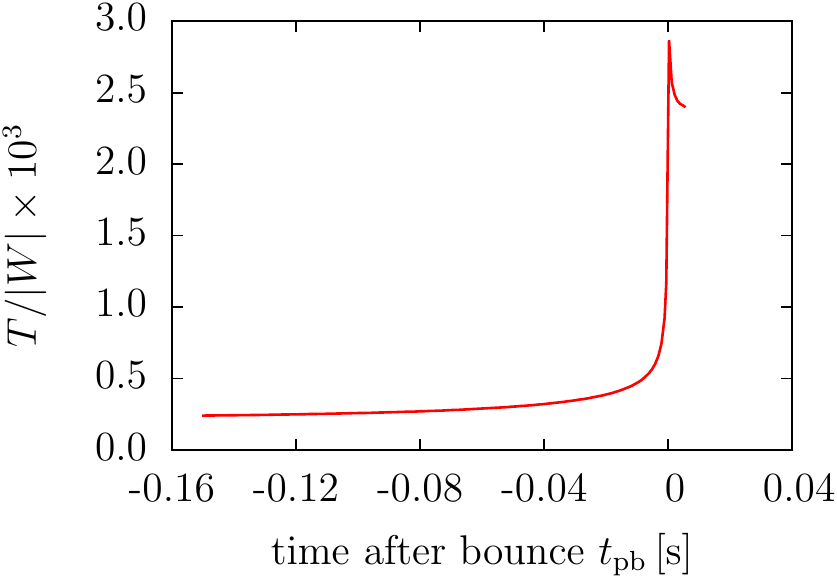}
\caption{\label{fig:toverw} Time evolution of $T/|W|$.}
\end{figure}

As a gauge of the degree of rotation, we show the ratio of the rotational energy to the absolute value of the gravitational energy, $T/|W|$ from the onset of collapse to just after bounce in figure \ref{fig:toverw}. It is found that $T/|W|$ varies from $\sim 2.5\times 10^{-4}$ to $\sim 3\times 10^{-3}$ during this period.

\begin{figure}[tbph]
\centering
\includegraphics[width=0.5\hsize]{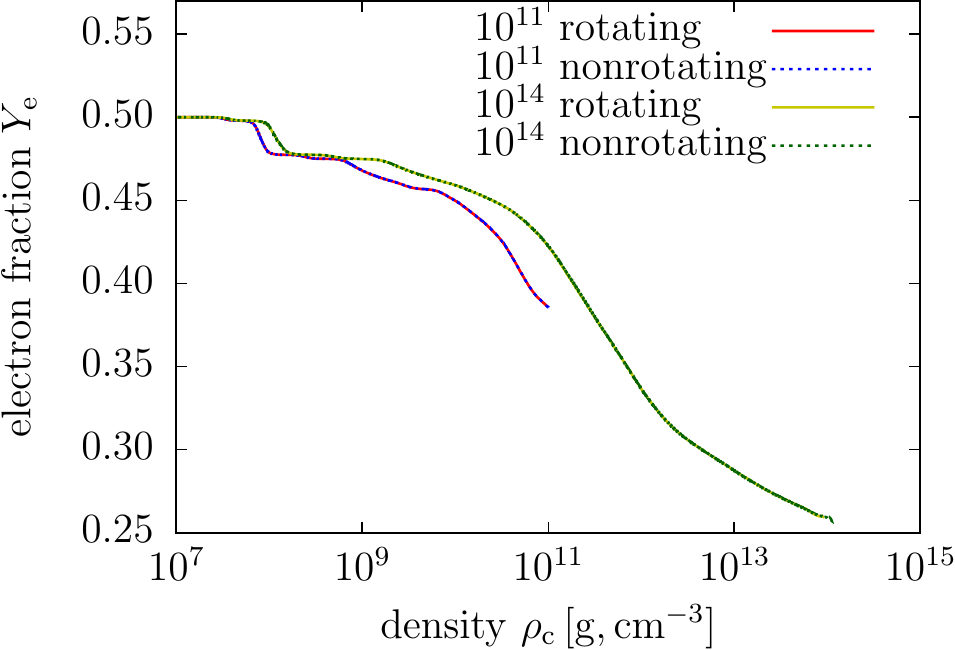}
\caption{\label{fig:yeprescription} Electron fraction profiles as a function of the density at the times when the central density is $10^{11}\,{\rm g\,cm^{-3}}$ (red and blue lines) and $10^{14}\,{\rm g\,cm^{-3}}$ (yellow and green lines). The solid and dotted lines represent our rotating model and the nonrotating model in \citet{2018ApJ...854..136N}, respectively.}
\end{figure}

\citet{2005ApJ...633.1042L} demonstrated in his 1D general relativistic Boltzmann-radiation-hydrodynamics simulations that the electron fraction $\ye$ of each fluid element follows approximately the same history during the collapse phase, which can be expressed conveniently as a function of density, whose functional form is obtained by fitting the numerical data. Note that his result is based on the 1D simulations, and possible effects of rotation on this ``$\ye$ prescription'' were not examined. Figure \ref{fig:yeprescription} shows the comparison between our rotating model and the nonrotating model in \citet{2018ApJ...854..136N}. It is clear that, contrary to the claim by \citet{2005ApJ...633.1042L}, the electron fraction profiles at different times cannot be expressed by a single function of density alone. This is not unexpected, though, since we use the updated electron-capture rates and assume the Newtonian gravity. There is almost no difference between the rotating and nonrotating models, on the other hand. The rotation assumed in this study is simply too modest, and more rapid rotation may change the result. Such investigations are currently being undertaken, and the results will be presented elsewhere in the near future.

\begin{figure}[tbph]
\centering
\includegraphics[width=0.5\hsize]{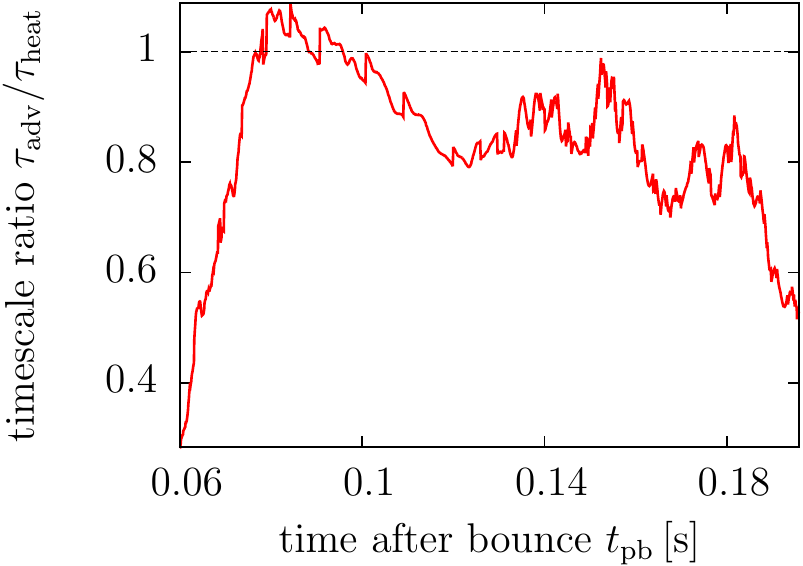}
\caption{\label{fig:timescale} Time evolution of the timescale ratio $\tau_{\rm adv}/\tau_{\rm heat}$.}
\end{figure}

Figure \ref{fig:timescale} shows the timescale ratio $\tau_{\rm adv}/\tau_{\rm heat}$, which is often used by supernova modelers. The advection timescale $\tau_{\rm adv}$ is defined as the ratio of the gain mass, which is the mass in the region where neutrino heating dominates cooling, to the mass accretion rate. The heating timescale $\tau_{\rm heat}$ is defined as $\tau_{\rm heat} = |E_{\rm gain}|/Q$, in which
\begin{equation}
E_{\rm gain} = \int_{r_{\rm gain}}^{r_{\rm shock}} \left(e_{\rm th} + \frac{1}{2}\rho v^2 + \rho \psi \right) \rd V
\end{equation}
is the total energy in the gain region, with $e_{\rm th}$ and $\psi$ being the thermal energy and gravitational potential, respectively, whereas $Q$ is the neutrino heating rate in the gain region. According to Appendix A in \citet{2016ApJ...818..123B}, the thermal energy should be defined as
\begin{equation}
e_{\rm th} = \frac{3}{2} \frac{\rho}{\bar{A}m_{\rm u}} kT + a T^4 + \left(e_{\rm e^\mp} - \ye m_{\rm e}c^2 \frac{\rho}{m_{\rm u}}\right),
\end{equation}
where $e_{\rm e^\mp}$, $\bar{A}$, and $a$ are the internal energy density of the electron--positron gas with their rest mass included, the mean mass number, and the radiation constant, respectively. When the ratio $\tau_{\rm adv}/\tau_{\rm heat}$ exceeds unity, the heating occurs faster than the advection and the supernova has a chance to explode successfully. It is seen in figure \ref{fig:timescale} that this happens during only a very short period and the ratio has decreasing trends thereafter, indicating the failure of shock revival in this model.

\begin{figure}[tbph]
\centering
\includegraphics[width=0.5\hsize]{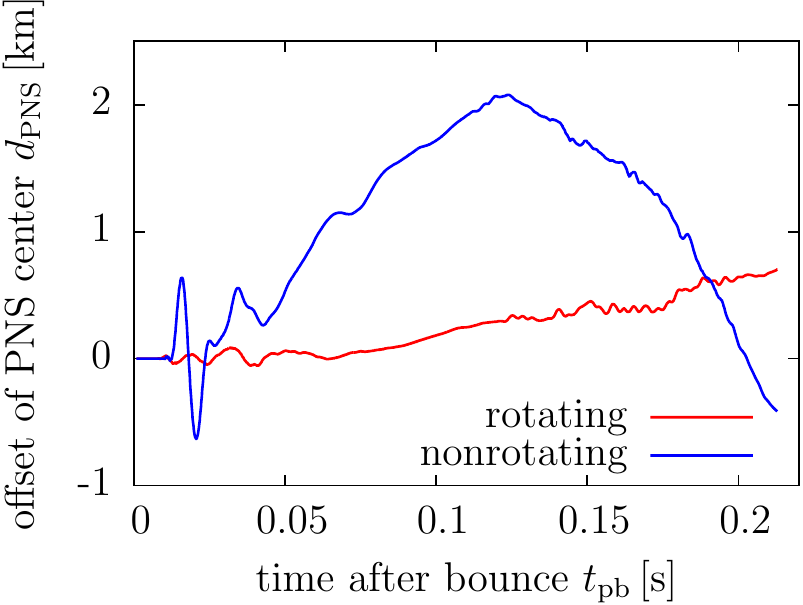}
\caption{\label{fig:pnskick} Comparison of the trajectory of the PNS center between the rotating (red line) and nonrotating (blue line; simulated in \citet{2018ApJ...854..136N}) models. The offset is measured in the laboratory frame.}
\end{figure}

Since our code is equipped with the moving-mesh capability, we can follow the proper motion of a PNS, unlike other codes, in which the center of a PNS is artificially fixed. This is shown in figure \ref{fig:pnskick} for both the rotating model presented in this paper and the nonrotating model presented in \citet{2018ApJ...854..136N}. According to the figure, the motion of the PNS is more violent in the nonrotating model than in the rotating model. This can be understood from figure \ref{fig:snapshots}. The entropy distributions in the meridian section are more symmetric with respect to the equator up to the stalled-shock phase in the rotating model. Since it is a result of the centrifugal force, the larger force imbalance between the northern and southern hemispheres leads to the more violent PNS kick in the nonrotating model. The kick velocity is small in both models, however, and the difference between the laboratory frame and the acceleration frame is also small accordingly.

\bibliographystyle{aasjournal}
\bibliography{ref}

\end{document}